\documentclass{emulateapj}
\shorttitle{GMCs in M64}
\shortauthors{Rosolowsky \& Blitz}
\begin{document}
\title{Giant Molecular Clouds in M64}

\newcommand{\co}{\ensuremath{\mbox{CO}}}
\newcommand{\cotope}{\ensuremath{^{13}\mbox{CO}}}

\author{E. Rosolowsky and L. Blitz}
\affil{Radio Astronomy Laboratory, 
University of California, Berkeley, CA 94720}
\email{eros@astro.berkeley.edu}

\begin{abstract}
We investigate the properties of Giant Molecular Clouds (GMCs) in the
molecule-rich galaxy M64 (NGC 4826).  In M64, the mean surface density
of molecular gas is $2N(\mbox{H}_2)\sim 10^{22}$ over a 2 kpc region,
equal to the surface densities of individual GMCs in the Milky Way.
We observed the $J=1\to 0$ transitions of \co, \cotope, and HCN.  The
line ratio $W_{\mathrm{CO}}/W_{\mathrm{13CO}}$ for $200\mbox{ pc} <
R_{gal} < 800\mbox{ pc}$ is comparable to that found in the Milky Way
and increases significantly outside this region, in part due to a
large contribution to the \co\ emission from diffuse gas, which
composes 25\% of the molecular mass in the galaxy.  We developed a
modified CLUMPFIND algorithm to decompose the \cotope\ emission into
25 resolved clouds.  The clouds have a luminosity--linewidth
relationship $L\propto\Delta V^{2.2\pm 0.4}$, substantially different
from the Milky Way trend reported by \citet{srby87}: $L\propto \Delta
V^5$.  Similarly, the clouds have a linewidth--size relationship of
$\Delta V\propto R_e^{1.0\pm 0.3}$ compared to $\Delta V\propto
R_e^{0.5}$ in the Milky Way. Estimates of the kinetic and binding
energies of the clouds suggest that the clouds are self-gravitating
and significantly overpressured with respect to the remainder of the
ISM in M64.  The \cotope-to-H$_2$ conversion factor is comparable to
what is seen in the Galaxy.  The M64 clouds have a mean surface
density at least 2.5 times larger than observed in Local Group GMCs,
and the surface density is not independent of mass as it is in the
Local Group: $\Sigma_{\mathrm{H2}}\propto M^{0.7\pm 0.2}$. The clouds
are correlated with the recombination emission from the galaxy,
implying that they are star forming; the rate is comparable to that in
other galaxies despite the increased densities of the clouds.  The
gas-to-dust ratio is similar to the Galactic value but the low
extinction in the visual band requires that the molecular gas be
clumpy on small scales.  We note that the internal pressures of clouds
in several galaxies scales with the external pressure exerted on the
clouds by the ambient ISM: $P_{int}\propto P_{ext}^{0.75\pm 0.05}$.
We attribute the differences between M64 molecular cloud and those in
the Local Group to the high ambient pressures and large molecular gas
content found in M64.
\end{abstract}
\section{Introduction}

Over 80\% molecular gas in the Milky Way is found in Giant Molecular
Clouds (GMCs) with masses $10^{3}~M_{\odot} <
M_{GMC}<10^{6.5}~M_{\odot}$ and sizes between 2 and 100 pc.  They have
a column density of $2N(\mbox{H}_2)\sim 10^{22}~\mbox{cm}^{-2}$,
independent of mass \citep{psp3}.  Observations of molecular gas
across the Local Group have found that most of the molecular gas in
these systems is also found in GMCs with properties similar to those
in the disk of the Milky Way \citep{nanten,eprb03,young-n205,smcgmcs}.
Beyond the Local Group, observers find several galaxies where the
typical molecular column density through the galactic disks is $>
10^{22}\mbox { cm}^{-2}$ \citep{song2} on kiloparsec scales implying
that the ISM is overwhelmingly molecular, since the column densities
of atomic gas are only rarely seen to exceed $10^{21}$~cm$^{-2}$.
Since this is larger than the typical column density through a single
GMC in the Local Group, the molecular ISM must be qualitatively
different in these galaxies.  Do the molecular clouds in these
galaxies blend together into a continuous sea of molecular gas or are
there discrete clouds of molecular gas that are denser than Local
Group GMCs?  Moreover, what are the star forming structures in this
molecular gas?  In the Local Group, star formation is predominantly
found in GMCs, but if there are no discrete clouds in this molecular
gas, do stars form uniformly throughout the galaxy from stellar mass
clumps in the sea of gas?  If there are GMCs but they are smaller and
denser than Local Group clouds, is the star formation efficiency
enhanced?  In order to answer these questions, we have conducted a
study of the molecule-rich galaxy M64 (NGC 4826) at high resolution to
search for GMC analogues in a molecule-rich environment.

M64 is a good target for such a study because it has a mean column
density across the central 2 kpc of the galaxy of $2N(\mbox{H}_2)=
10^{22}~\mbox{cm}^{-2}$, making it an intermediate case between the
Local Group and starburst galaxies.  The galaxy is one of the closest
molecule rich galaxies \citep[4.1~Mpc,][]{ngc-tully}.  A merger with a
small, gas-rich galaxy roughly 1 Gyr ago is thought to be responsible
for the relatively large molecular gas content for its early Hubble
type (SA) \citep[][B94]{counterrot_m64}.  The 3$\farcs$5 resolution
obtained with the BIMA millimeter interferometer projects to 75 pc,
which would be sufficient to resolve the largest molecular clouds in
the Milky Way disk at the distance of M64.  Two other recent studies
examine properties of molecular emission in the nucleus of M64.
\citet[][ M02]{meier-thesis} conducted a study of the nuclear region
using the OVRO millimeter array in the 3 mm transitions of \co,
\cotope, and C$^{18}$O.  The primary beam at OVRO has a FWHM of 65$''$
which samples gas to $R_{gal}=600$ pc along the major axis and the map
has a $5''$ synthesized beam.  These observations were principally
used to determine the bulk properties of the molecular gas using LVG
analyses.  \citet[][ hereafter NUGA]{NUGA} also conducted an extensive
analysis of the molecular gas in the nuclear region using the IRAM
Plateau de Bure interferometer (PdBI) in the \co$(2\to 1)$ and the
\co$(1\to 0)$ transitions.  With the 42$''$ primary beam, the PdBI
observations cover the galaxy to $R_{gal}=420$ pc for the $(1\to 0)$
transition with a $2.5''$ resolution. The NUGA project focused
interpretation of their data on the mechanisms by which the central
AGN could be fed.  Our search for molecular clouds in M64 is
complementary to the aims of both M02 and NUGA.  With a 105$''$
primary beam and $3.5''$, our BIMA study maps the galaxy to the edge
of the molecular disk at $R_{gal} \gtrsim 1000$~pc permitting a more
thorough study of the molecular gas.  In addition, this study includes
zero-spacing observations of \co\ and \cotope\ which are essential in
searching for low surface brightness gas among the clouds.  In this
paper, we report on our observations (\S\ref{observations}), examine
the dynamics of the molecular disk (\S\ref{rc}) and variations in the
isotopic line ratios (\S\ref{line_ratios}), decompose the blended
emission into cloud candidates and argue that these candidates
represent the high density analogues of Local Group GMCs
(\S\ref{decomp}).

\section{Observations}
\label{observations}
\subsection{Molecular Line Observations}
We observed M64 using the BIMA array \citep{bima} in three tracers of
molecular gas: \co$(1\to 0)$, \cotope$(1\to 0)$ and HCN$(1\to 0)$.  We
utilized data from the B, C and D arrays, including medium-resolution
(C array) and zero-spacing (UASO 12-m) observations made in the BIMA
Survey of Nearby Galaxies \citep[SONG,][]{song2}.  The combined
observations resulted in a synthesized beam of $3.5''$ which projects
to a linear scale of 75 pc.  Details for the individual observations
appear in Table \ref{obsprops}.  We calibrated the visibility data
using the same techniques as employed in SONG.  The visibility data
were inverted using natural weighting to maximize signal-to-noise
ratios.  The high resolution \co\ observations had slightly different
pointing centers with respect to the low resolution and zero-spacing
data, requiring inversion to the image domain of the high and low
resolution data separately.  We then combined these CO maps using the
linear combination method \citep{lincom,song2}.  Images were then
cleaned using a hybrid H\"ogbom / Clark / SDI algorithm with the
residuals of the cleaned components added back in to avoid any flux
loss.  To check for possible contamination by continuum emission from
the AGN, we imaged the LSB data from \cotope\ and HCN. The lack of
sources in the resulting maps places a 3$\sigma$ point source
sensitivity of 3.4 mJy at 85 GHz for objects in the galaxy and a 3.2
mJy limit at 106 GHz.

\begin{deluxetable*}{cccccc}
\tablecaption{\label{obsprops}Properties of BIMA Array Observations}
\tablewidth{0pt}
\tablehead{
\colhead{Tracer} & \colhead{Array} & \colhead{Length}
& \colhead{$\langle T_{sys}\rangle$} & \colhead{Beam Size} & 
	\colhead{$\sigma_{rms}$\tablenotemark{a}}\\
& &  [hours] &  [K] & [$''$] & [Jy bm$^{-1}$ km s$^{-1}$]
}
\startdata
$^{12}$CO$(1\to 0)$ & B & 11 h 20 m & 610 & $4.6 \times 3.6$ & 0.24 \\
& C\tablenotemark{b} & 10 h 57 m & 417 & & \\

$^{13}$CO$(1\to 0)$ & B & 22 h 35 m & 264 & $3.3 \times 2.6$ & 0.11 \\
& C & 12 h 05 m & 430 & &\\
& D & 3 h 35 m & 305 & &\\
HCN$(1\to 0)$ & B & 16 h 01 m & 198 & $4.0 \times 3.4 $ & 0.037 \\
& C & 7 h 47 m & 221 & & \\
& D & 3 h 58 m & 194 & & \\
\enddata
\tablenotetext{a}{Using a 4.25 km s$^{-1}$ channel.}
\tablenotetext{b}{C array data in $^{12}$CO($1\to 0$) are from the BIMA
Survey of Nearby Galaxies \citep{song}.}
\end{deluxetable*}

\begin{figure}
\begin{center}
\plotone{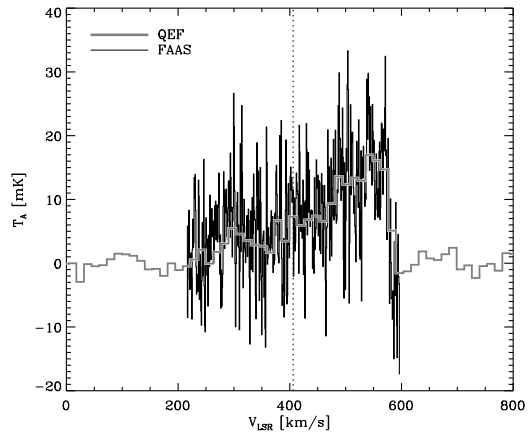}
\caption{\label{fcrao_spec} Example of resulting spectrum from
stitching procedure.  This spectrum is taken at 
$\alpha_{2000}=12^{\mathrm{h}}56^{\mathrm{m}} 42\fs7,
\delta_{2000}=+21^{\circ} 41' 18\farcs 5$
 spectra are merged around the LSR velocity
of the galaxy, shown with the dotted line at 406 km s$^{-1}$.
This shows the general success of producing a combined high-resolution
spectrum from different velocity configurations.}
\end{center}
\epsscale{1.0}
\end{figure}

Although M64 has been partially mapped with the IRAM 30-m telescope
\citep{iram_m64}, the data are no longer available in an electronic
format (F. Casoli, private communication).  To provide a fully-sampled
\cotope\ map, we observed the galaxy with the FCRAO 14-m telescope.
We observed 7 points arranged in a hexagonal pattern in order to
obtain a Nyquist-sampled map and recover all spatial frequencies.
Observations of SiO maser stars through the night were used to check
the pointing solution of the telescope.  We observed the galaxy by
alternately pointing at the galaxy with two different pixels in the
SEQUOIA 32-pixel receiver.  We used the time that each pixel was
pointed at blank sky as the reference position for that pixel.  Vane
calibration was used to set the flux scale.  We used two back ends in
parallel: the Quabbin Extragalactic Filterbank (QEF), which provided
320 MHz of spectral coverage in 64 5-MHz channels (864 km s$^{-1}$
divided into 13.5 km s$^{-1}$ channels), and the Focal Plane Array
Autocorrelation Spectrometer (FAAS), which provided 80 MHz of spectral
coverage with 0.3125 MHz channels (216 km s$^{-1}$ divided into 0.84
km s$^{-1}$ channels).  Unfortunately, the velocity range of 216 km
s$^{-1}$ provided by the FAAS correlator is insufficient to cover the
entire galaxy and the velocity resolution of the QEF was too coarse to
merge with the interferometer data.  Therefore, we had to observe the
galaxy with two separate velocity configurations which measured the
approaching and receding halves of the galaxy separately.  We used the
QEF data to establish the common scaling between the two velocity
configurations.  Since the FAAS and QEF follow the same amplifier
chain, smoothing FAAS data to the QEF resolution should result in the
same spectrum.  To stitch together the two sets of FAAS data, we
subtracted a linear baseline from the QEF data and used the resulting
spectrum as a model of signal emission in the FAAS data.  By
accounting for the expected signal using the QEF data, a linear
baseline could be removed from each of the FAAS spectra separately and
spectra from the two configurations could be merged.  An example of
this appears in Figure \ref{fcrao_spec}.  Typically, the spectra had a
noise level of 12 mK in a 4.25 km s$^{-1}$ channel after correcting
for a main beam efficiency of 0.5 at 110 GHz.  As a check, we smoothed
the final FAAS spectrum to the resolution of the QEF data and found
there were only small variations resulting from higher order baselines
in the data sets.

We merged the zero-spacing information with the \cotope\
interferometer data using the Linear Combination technique
\citep{lincom,song2}.  The merged dataset was deconvolved using a
CLEAN algorithm which terminated at the 1.5$\sigma_{rms}$ level of the
data.  The combined dataset contains $94\pm 7 \%$ of the single dish
flux.  Finally, we regridded the \co\ and HCN images to match the
spatial and velocity scale of the \cotope\ data to facilitate
pixel-to-pixel comparison, resulting in a final channel width of 4.25
km s$^{-1}$ in all maps.

\begin{deluxetable*}{ccccc}
\tablecaption{Flux Recovery for BIMA Observations\label{fluxtable}}
\tablewidth{0pt}
\tablehead{
\colhead{Tracer} & \colhead{Flux} & \colhead{SD Flux}
& \colhead{Frac.}  & \colhead{Completeness Lmt.\tablenotemark{c}}\\
 & Jy km s$^{-1}$& Jy km s$^{-1}$ &  Recovered & Jy km s$^{-1}$
}
\startdata
\co\ & $1678\pm 23$ & $1845 \pm 217$\tablenotemark{a} & 
$91\pm 12$\% & 1.06 \\
\cotope\ & $160 \pm 6$ & $230 \pm 10 $ & $70\pm 6$\% & 0.35 \\
HCN & $35 \pm 3$ & $36.5\pm 5.6$\tablenotemark{b} 
& $96\pm 18$\%& 0.20\\
\enddata
\tablenotetext{a}{\citet{song2} }
\tablenotetext{b}{\citet{hcn-helfer}}
\tablenotetext{c}{For recovery of 90\% of point sources with velocity
FWHM of 8.5 km s$^{-1}$ (two channels).}
\end{deluxetable*}

\subsection{Recombination Line Images}
\label{recomb}
To obtain recombination line emission images, we downloaded HST images
of the galaxy in Pa$\alpha$ (F187N), $H$-band continuum (F160W),
H$\alpha$ (F656N) and $V$-band continuum (F547M) from the HST data
archive.  The Pa$\alpha$ image was presented originally by
\citet{nicmos-atlas}.  We followed the reduction steps presented in
that work including removal of the `pedestal effect' in the NIC3
camera.  However, we adopted a different optimal scaling between the
F160W continuum image ($W_{ij}$) and the F187N line image ($N_{ij}$)
for continuum subtraction.  We minimize a
goodness-of-fit parameter $G(m)=\sum_{ij}|(N_{ij}-m
W_{ij})/\sigma_{ij}|^p$ with respect to the scaling $m$, where
$\sigma_{ij}$ is the noise associated with each pixel and the power
$p$ is determined by
\[p=\left\{\begin{array}{cl} 0.5, & N_{ij} \geq m W_{ij} \\
6, & N_{ij} < m W_{ij}. \end{array}\right.\] 
This construction preferentially fits those points in the narrow band
image that are free from Pa$\alpha$ emission.  For the NICMOS data,
$m=0.0418$.  This method implicitly assumes that the color of the
continuum does not vary across the map.

The H$\alpha$ images have not been published, and were observed under
HST GO Proposal \#8591 (Richstone, D.).  We used a similar procedure
to that of the Pa$\alpha$ analysis to extract an image containing only
H$\alpha$ line flux from the archival data.  The principal differences
were that we used median filtering among an array of images to
eliminate cosmic rays.  We used a F547M image of the galaxy for the
continuum image and the F656N image containing the line flux.  The
optimal scaling between these two images is, on average, $m=0.304$
though data from each chip of the WFPC2 were treated separately.

\section{Analysis of the Molecular Line Maps}
\label{linemaps}
\begin{figure*}
\begin{center}
\plotone{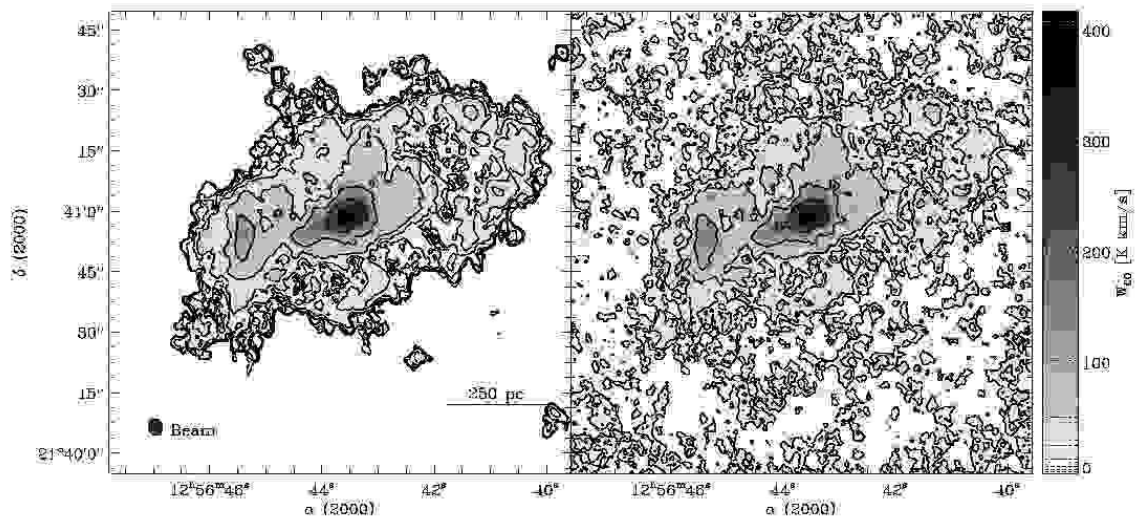}
\caption{\label{comap} Masked integrated emission map of M64 in
$^{12}$CO$(1\to 0)$ for two different masking methods.  The masking
method used in the paper (left panel) is compared to a mask generated
by clipping the emission above $2\sigma_{rms}$ (right panel).
Comparing the two methods highlights the ability for the adopted
method to recover low surface brightness emission.  In the left panel,
contours are $2^k\sigma_{rms}$ for $k \ge 0$ and in the right panel,
contours are $2^j\sigma_{rms}$ for $j\ge 3$.  .  The levels of the
contours relative to the grayscale are indicated in the colorbar by
dotted lines.}
\end{center}
\end{figure*}

\begin{figure}
\begin{center}
\plotone{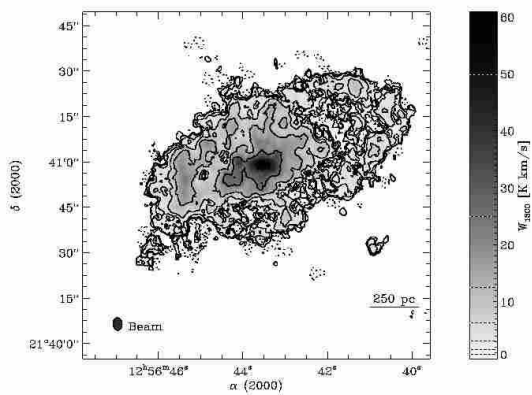}
\caption{\label{cotopemap} Masked integrated emission map of M64 in
$^{13}$CO$(1\to 0)$.  The contours are $-2\sigma_{rms}$,
$-1\sigma_{rms}$ (dotted) and $2^k\sigma_{rms}$ for $k \ge 0$
(solid).  The levels of the
contours relative to the grayscale are indicated in the colorbar by
dotted lines. The emission appears significantly more clumpy than the
$^{12}$CO emission despite similar $UV$ coverage.}
\end{center}
\end{figure}

\begin{figure}
\begin{center}
\plotone{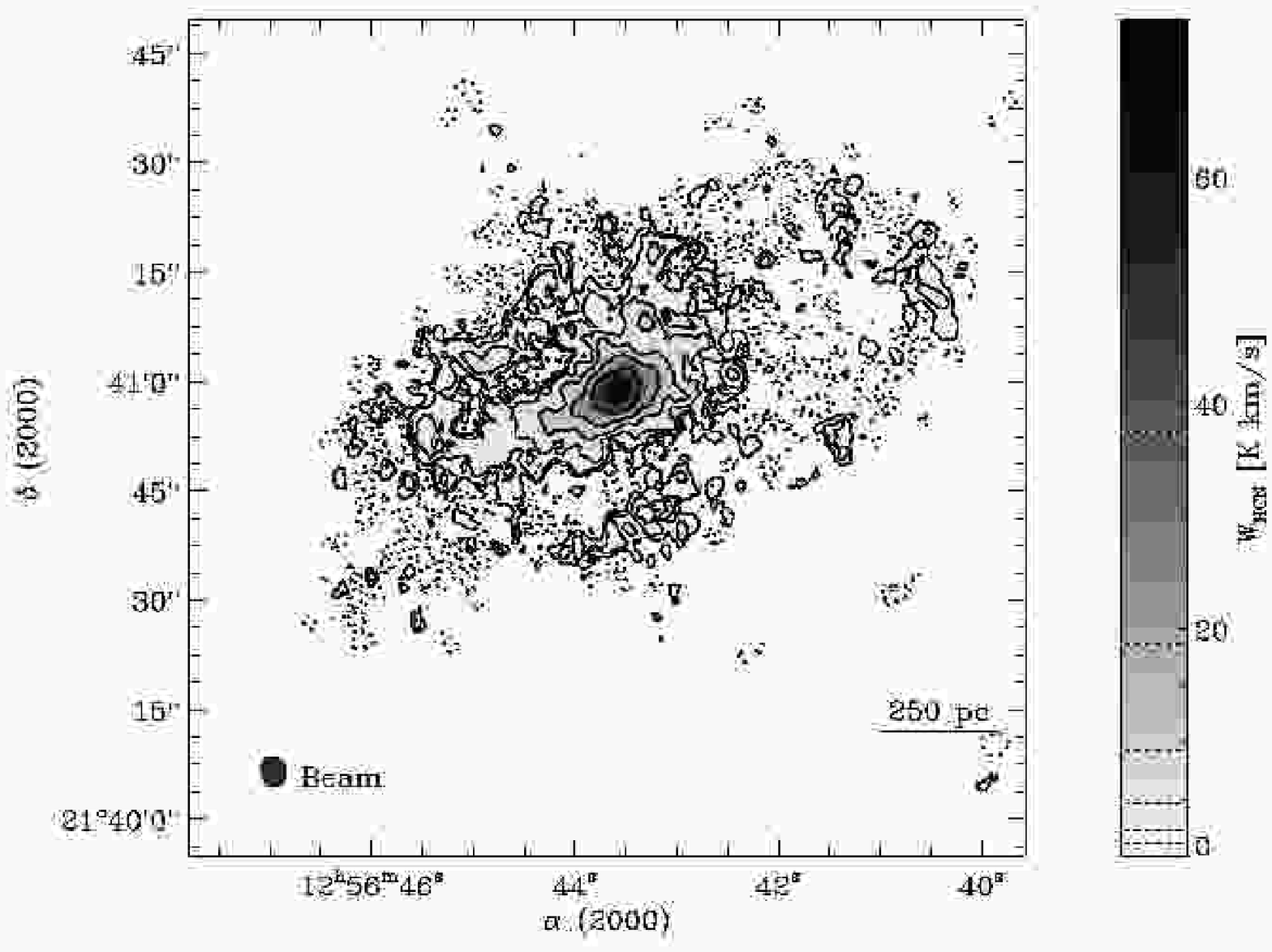}
\caption{\label{hcnmap} Masked integrated emission map of M64 in
HCN$(1-0)$.  The contours are $-2\sigma_{rms}$, $-1\sigma_{rms}$
(dotted) and $2^k\sigma_{rms}$ for $k \ge 0$ (solid).  The levels of
the contours relative to the grayscale are indicated in the colorbar
by dotted lines.  The emission is strongly peaked in the nuclear
region of the galaxy, but the masking procedure highlights faint
emission, particularly in the NE half of the disk.}
\end{center}
\end{figure}

In Figures \ref{comap} to \ref{hcnmap}, we present velocity-integrated
maps of the molecular disk in M64.  The data have been masked to
highlight the structure of the emission in the three maps.  Since all
three data sets have the same coordinate axes, a single mask is
generated common to all three data sets.  An element is included in
the mask if it has detectable emission in CO.  For the CO data, the
masking process begins by searching for all positions with emission
($I$) greater than 4 times the rms value of the noise ($\sigma_{rms}$)
in {\it two adjacent} velocity channels.  These regions are then
expanded to include all pixels with $I \ge 2\sigma_{rms}$ that are
connected in position or velocity space by $\ge 2\sigma_{rms}$
detections to the joint $4\sigma_{rms}$ detection ``core'' following
the method used in \citet{eprb03}.  For comparison, we have plotted an
integrated intensity image using a 2$\sigma_{rms}$ clip in the
right-hand panel of Figure \ref{comap}.  The mask generated from the
CO is then applied to the \cotope\ and HCN emission.  Using a common
mask among the different tracers facilitates direct comparison of the
morphology of the molecular emission.  We use this process since
emission from rare tracers will be located in the regions of the
datacube where there is emission from the more common tracers.
Because of this assumption, the masked maps have complex statistical
properties since the CO map contains only positive emission while the
\cotope\ and HCN emission include noise in the data.  Nonetheless, the
masking process highlights faint HCN emission in the disk of the
galaxy which is difficult to detect by analyzing the HCN emission
alone (Figure \ref{hcnmap}).  These faint features align with the
strong CO features seen in Figures \ref{comap} and \ref{cotopemap}.
We used Monte Carlo simulations to establish point source completeness
limits with this masking algorithm for emission with a velocity FWHM
of 2 channels.  The quoted limits in Table \ref{fluxtable} are the
threshold for a point source to be recovered in 90\% of the Monte
Carlo trials.  We use this masking process to examine the structure of
the emission and for the rotation curve but use simpler methods for
line ratio studies and cloud decomposition.

\begin{figure}
\begin{center}
\plotone{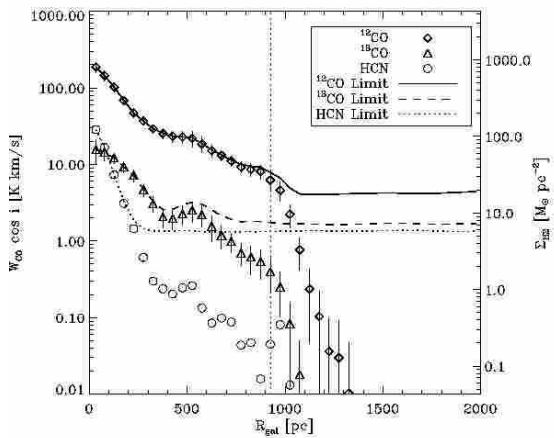}
\caption{\label{surfdens} Azimuthally averaged surface brightness for
molecular tracers.  The averages were taken in 50 pc bins.  The error
bars represent 1$\sigma$ errors and the lines indicating upper-limits
represent the profile that would be observed if the blank pixels
contained emission at the $3\sigma_{rms}$ level, a very conservative
upper limit.  The averages have been scaled by $\cos i = \cos
56^{\circ}$ to reflect face on values.  The \co\ emission shows a
marked truncation at $R_{gal}=925\pm 25$ pc, indicated by the vertical
dotted line.}
\end{center}
\end{figure}

The three maps share several common features.  The emission is
strongly peaked at the center of the galaxy with a surrounding
disk. The NE half of the disk shows stronger emission than the SW
half.  The molecular emission in the NE half of the galaxy aligns with
the prominent dust lane that earns the galaxy's nickname of the `Evil
Eye;' however, these observations show that there is significant
molecular gas in the SW of the galaxy as well.  The disk appears to
cut off sharply at a galactocentric radius of $R_{gal}=1000$ pc.  To
quantify this, we plot the azimuthally averaged surface brightness
distributions in Figure \ref{surfdens}.  The azimuthal average shows a
significant cutoff at $R_{gal} = 925 \pm 25$ pc in the \co\ emission,
well above the $3\sigma_{rms}$ noise level of the observations.  A similar
cutoff is also seen in the \cotope\ emission and likely exists in the
HCN emission but is below our significance threshold.  

\subsection{The Rotation Curve of M64}
\label{rc}

M64 is peculiar because the outer \ion{H}{1} disk ($R_{gal} > 2$ kpc)
is counterrotating with respect to the inner \ion{H}{1} and stellar
disks.  Studies by \citet{counterrot_m64} and
\citet{m64_stellar} suggest that the outer disk is result of a
retrograde merger with a small galaxy.  They propose that interaction
between the counterrotating gas disk and the prograde stellar+gas disk
dissipates significant angular momentum, fueling infall to the central
disk which accounts for the high values of the surface mass density of
neutral gas \citep{m64_stellar,m64_rubin}.  However, the gas in the
central region of the galaxy ($R_{gal} \lesssim 1$~kpc) appears to be
in a normal, rotating disk which is decoupled from the peculiar
dynamics of the outer galaxy.  The outer edge of the molecular disk
occurs at the interface between the well-organized inner disk and the
infalling ionized gas in the in region $1000 \mbox{ pc}< R_{gal}< 1500
\mbox{ pc}$.

Since these are the first millimeter observations to cover the
entirety of the molecular gas disk, we examined the dynamics of the
inner region of the galaxy.  We generated velocity fields from the
masked emission for each of the three tracers.  We analyzed the
velocity fields using the tilted-ring algorithm first presented by
\citet{rotcur} to generate a rotation curve as implemented in the NEMO
stellar dynamics toolkit \citep{nemo}.  We used literature values for
initial estimates of the circular rotation speed, position angle,
inclination, systemic velocity and dynamical center (B94, NUGA).  We
then iteratively fit for these parameters until a convergent model was
determined following the procedure described in \citet{wbb04}.  The
derived parameters of the disk do not vary significantly for the
molecular emission within $40''$ of the galactic center, so we forced
the inclination, position, systemic velocity, and position angle of
the rings to be equal in order to improve the estimate of the
dynamical properties.  The dynamical center of the galaxy is at
$\alpha_{2000}= 12^{\mbox{\scriptsize h}}\ 56^{\mbox{\scriptsize m}}\
43\fs 6$ and $\delta_{2000}=21\degr\ 40\arcmin\ 57\farcs 7$ with a
systemic velocity of $V_{sys}=411\pm 1 \mbox{ km s}^{-1}$.  The disk
of the galaxy has a position angle of $-67\degr\pm 3\degr$ and an
inclination $i=59\degr\pm 2\degr$ averaged over $R_{gal}<40''$ using
4$''$ bins including corrections in the quoted errors for beam
smearing and oversampling.  These dynamical properties agree well with
those adopted by other authors.  The peak of the molecular emission is
slightly offset ($1''$) from the dynamical center of the galaxy, as is
found in many disk galaxies \citep{song2}.

\begin{figure}
\begin{center}
\plotone{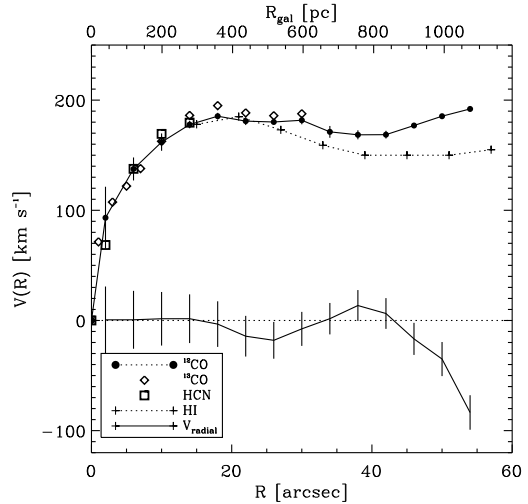}
\caption{\label{rotcur} Rotation curve of M64.  The data agree well
until the small numbers of points that show $^{13}$CO emission lead to
poor fits in the outer section of the galaxy.  Typical errors in the
derived rotation velocities vary between 0.5 and 10 km s$^{-1}$ and
$1\sigma$ error bars are plotted for the $^{12}$CO points.  We also
plot the radial velocity derived from an axisymmetric inflow/outflow
model.}
\end{center}
\end{figure}

The rotation curve of the galaxy is shown in Figure \ref{rotcur} for
each of the three tracers and the \ion{H}{1} (from B94).  The rotation
curves agree well for the three tracers in the inner portions of the
galaxy.  At large radii, the \ion{H}{1} rotation curve differs from
that of the molecular gas.  This may be a physical effect arising from
infalling material interacting with the molecular disk.  Radial
motions are seen in the atomic gas at larger radii ($R_{gal}\sim
80''$, 1.6 kpc, B94), so we fit the amplitude of an axisymmetric
radial gas flow from the velocity field of the galaxy, which is
plotted as $V_{radial}$ in Figure \ref{rotcur}.  We find that there
are significant radial motions of molecular gas for $R_{gal} >40''$
(800~pc), with amplitudes comparable to the motions found at the inner
edge of the counterrotating \ion{H}{1} disk ($V_{radial} = -70 \mbox{
km s}^{-1}$, B94).  In the region between $40'' < R_{gal} < 60''$, we
may be observing the interface between infalling gas and the rotating
molecular disk.  However, the molecular disk appears fairly asymmetric
beyond $R_{gal}=40''$ (800~pc) and radial motions may be the result of
elliptical gas orbits.  We also confirm the result of M02 who noted
that that the \co\ emission at the NE end of the minor axis is
significantly displaced in velocity from the circular rotation of the
galaxy.  Finally, we see the same evidence for streaming motions in
both \co\ and \cotope\ emission that is discussed in NUGA in the inner
parts of the molecular disk.

\subsection{Line Ratios}
\label{line_ratios}

The \co\ map (Figure \ref{comap}) appears to be significantly smoother
than that of the \cotope\ map (Figure \ref{cotopemap}).  Initially, it
is unclear whether this reflects differences in the actual emission
distribution or is due to noise because the dynamic range in the \co\
map is twice as large as that in the \cotope\ map.  However, as we
discuss below, the difference in the morphologies is real and arises
from variations in the fraction of molecular gas found in a diffuse,
translucent phase\footnote{Throughout the paper, ``diffuse'' gas
refers to molecular gas that is not self-gravitating and is used in
contrast with ``bound'' or ``self-gravitating.''  Diffuse gas is often
also ``translucent'' meaning $A_V \lesssim 3$ as opposed to bound gas
which is often ``opaque'' meaning $A_V \gtrsim 3$.}.

Because the abundance of \cotope\ is lower than that of \co\,
typically by a factor of 40 to 70 \citep{ism-abund}, the optical depth
of the \cotope\ line is significantly smaller than that of the \co\
line. In a simple radiative transfer model through a slab of
chemically uniform, isothermal gas, the line ratio $R_{13}\equiv
W_{\mathrm{12CO}} / W_{\mathrm{13CO}}$ should fluctuate between the
abundance ratio of the two species in the optically thin limit to
unity in the optically thick limit.  Averaged over a portion of the
Milky Way disk, $R_{13}=6.7$ \citep[$33.5^{\circ} < b
<35^{\circ}$,][]{mw-1213}.  However, the value of $R_{13}$ varies depending
on the structures observed.  Giant Molecular Clouds have $R_{13} = 4.5$
\citep{mw-1213} while $R_{13}=10\to 20$ in translucent, high-latitude
molecular clouds \citep{highlat,highlat2} and $R_{13}=21\pm 8$ is observed
in small molecular structures in the ISM \citep{tiny-clouds}.  At the
most fundamental level, the variations in $R_{13}$ reflect variations in
the relative opacities of the two line transitions.  Unfortunately,
radiative transfer in molecular gas remains a daunting problem with
solutions existing only for relatively simple problems.  Nonetheless,
since bound, opaque clouds show a systematically lower value of $R_{13}$
than do diffuse, translucent clouds, we view variations in $R_{13}$ as
empirical evidence for a changing fraction of material in bound
clouds.

In addition to having a higher value of $R_{13}$, translucent molecular
structures also exhibit a lower CO-to-H$_2$ conversion factor.  We
express the \co-to-H$_2$ conversion as
\[ N(\mbox{H}_2) = 2 \times 10^{20}\mbox{ cm}^{-2}\cdot X_2\cdot \left(
\frac{W_{\mathrm{12CO}}}{1 \mbox{ K km s}^{-1}}\right)\] where $X_2=1$
for massive clouds in the Milky Way \citep{sm96,dht01} and M33
\citep{rpeb03}.  In translucent molecular clouds, $X_2\sim 0.25$
\citep{xfac-translucent,xfac-ch}, i.e., \co\ emission from
translucent structures is {\it overluminous} relative to the molecular
column in translucent clouds when compared to more massive
structures. Since these regions also exhibit high values of $R_{13}$, the
measurements of $R_{13}$ can, therefore, be used to highlight regions where
the assumption of $X_2=1$ is questionable.

Variations in $R_{13}$ have been observed in extragalactic molecular
clouds.  \citet{m33-1213} used the Owens Valley interferometer to show
that $R_{13}=7.5$ in a GMC in M33; but when the GMC and the surrounding gas
was observed using a single dish telescope, they found $R_{13}=10$.  They
attributed the difference between the $R_{13}$ values on different scales
to the presence of diffuse molecular gas surrounding the GMC with $R_{13} >
13.5$.  Beyond the Local Group, \citet{cotope-ratio} find significant
variations in $R_{13}$ over the face of galaxies with increases in the
centers of galaxies.  \citet{starburst-1213} find that starbursting or
interacting galaxies have systematically higher values of $R_{13}$.  These
regions likely contain a larger fraction of diffuse molecular gas than
is seen in the Milky Way disk.

\begin{figure}
\begin{center}
\plotone{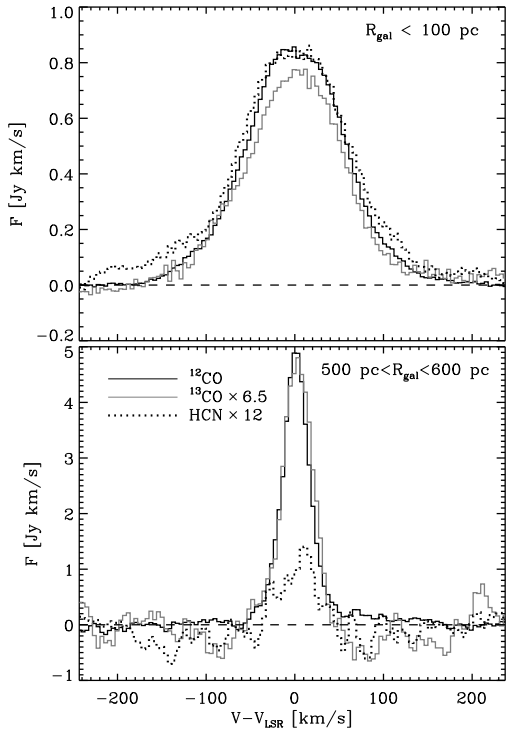}
\caption{\label{speccomp} Two examples of the spectra derived from the
averaging process used to measure line ratios.  The top panels shows
the emission averaged over $R_{gal} < 100$ pc and the bottom panel
shows the average for $500 \mbox{ pc} < R_{gal} < 600 \mbox{ pc}$.
Emission from \cotope\ has been scaled up by a factor of 6.5 and
emission from HCN has been scaled up by a factor of 12.  The line
ratios change significantly over the face of the galaxy.}
\end{center}
\end{figure}

We use the variation of $R_{13}$ across the face of M64 as proxy for
whether the emission appears to rise predominantly from highly opaque
objects ($R_{13} \sim 4$) or translucent objects ($R_{13} > 10$).  Since the
masking process likely introduces systematic effects into the
generation of moment maps, we use the unmasked data.  In order to
increase the significance of a measurement, we averaged together
several spectra from annuli in the plane of the galaxy.  We used the
rotation curve from \S \ref{rc} to shift each spectrum by the
projected rotation velocity at each position which places the emission
from each spectrum at a common velocity. The magnitude of the shift is
$-V(R_{gal}) \cos \theta \sin i$, where $\theta$ is the polar angle from the
kinematic major axis, measured in the plane of the galaxy, and $i$ is
the inclination.  This correction does not account for radial motions
in the galaxy which are significant at $R_{gal}>800$~pc.  This averaging
process is so effective that it permits the measurement of HCN
emission to large galactic radii, even though little emission can be
clearly identified in the data cubes.  Examples of the averaging
appear in Figure \ref{speccomp} for two annuli: $R_{gal} < 100$ pc and
$500 \mbox{ pc} < R_{gal} < 600 \mbox{ pc}$, which show substantially
different line ratios, particularly in HCN.  The spectrum for $R_{gal}
< 100$ shows evidence for the HCN/CO line ratio varying over the line
with the wings of the line showing comparatively more HCN emission than
the core of the line.  The wings of the line are generated from region
at $R_{gal}\sim 0$~pc where the velocity gradient across the synthesized
beam is the largest.  The larger line ratio in the wings suggests that
the molecular gas is densest in the region near the AGN (c.f., NUGA).

\begin{figure*}
\begin{center}
\plottwo{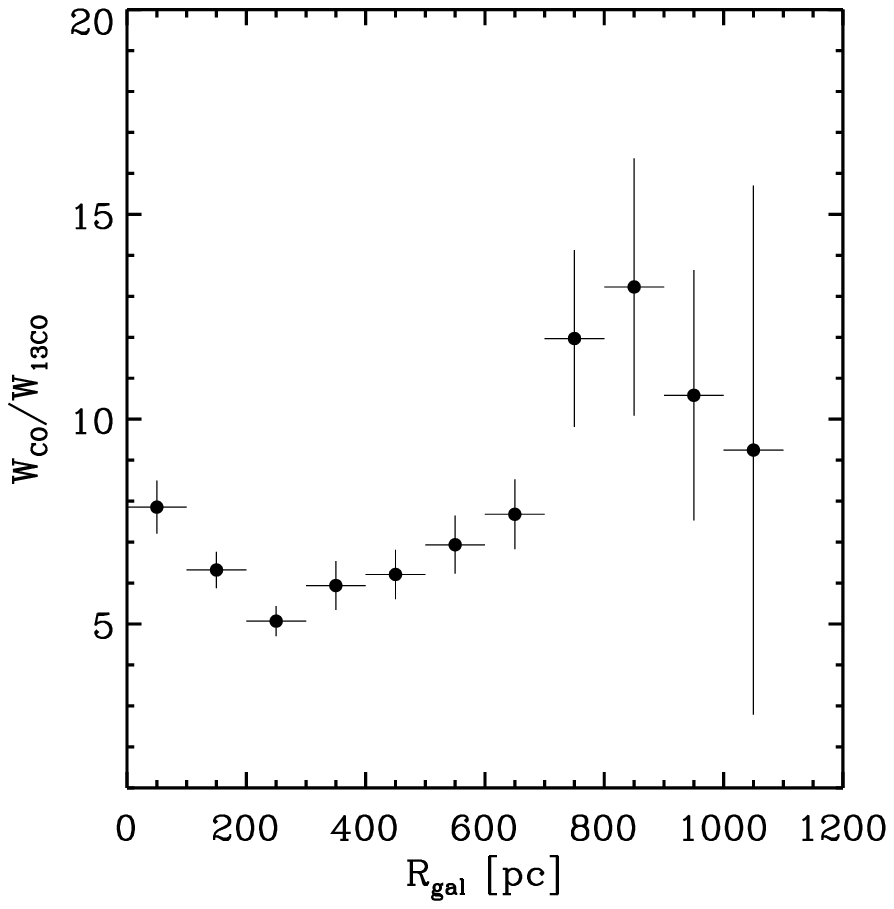}{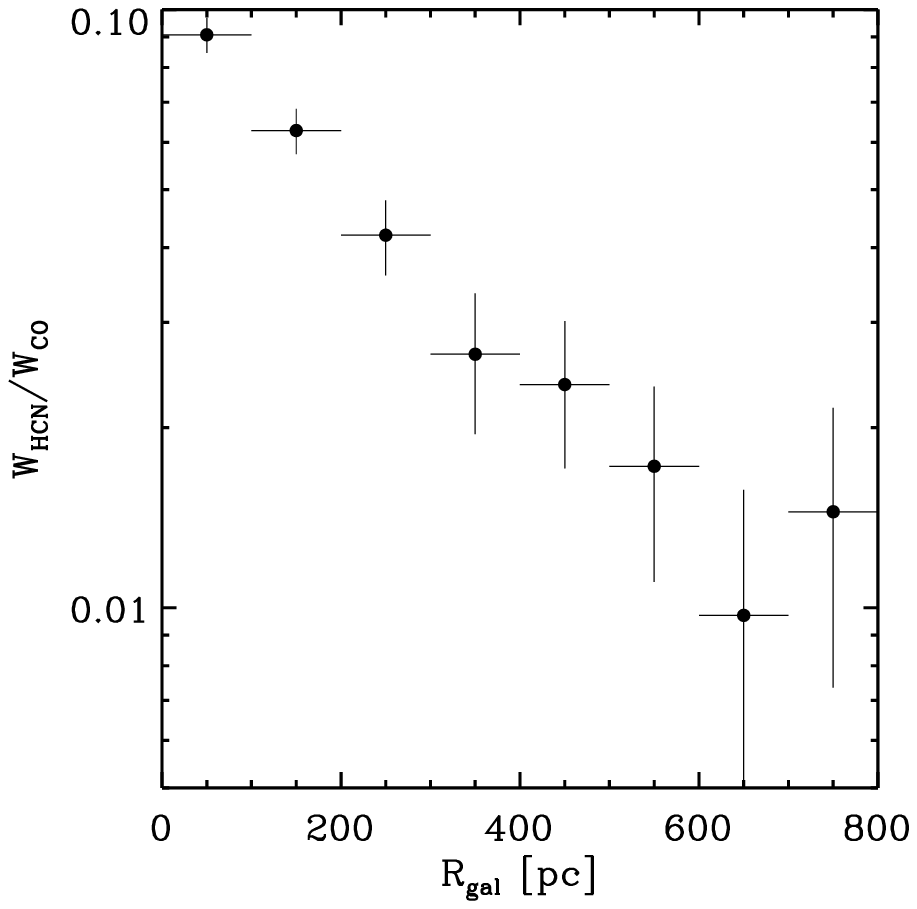}
\caption{\label{linerat} Line ratios between molecular gas tracers as
a function of galactocentric radius.  Over most of the disk, the \co\
to \cotope\ ratio has a values typical of the Milky Way disk ($\sim
6$) but shows a significant increase at small and large values of
$R_{gal}$.  For $R_{gal}>800$ pc, there are significant radial motions
which may introduce systematic errors into the measurement technique.
These increases may indicate significant amounts of diffuse molecular
gas.  The HCN/\co~ratio is larger than seen in the GMCs in the
Solar neighborhood, implying the average volume density of molecular
gas is higher in M64 than it is in the Milky Way.}  \epsscale{1.0}
\end{center}
\end{figure*}

We measured the line ratio as a function of position across the galaxy.
We integrated the emission over the derived spectra to measure the flux in
annuli of galactic radii 100 pc wide.  We then took ratios of the
fluxes and plot the results in Figure \ref{linerat}.
The value of $R_{13}$ changes significantly over the face of the
galaxy with most of the molecular disk having a value of $R_{13} \sim
6$ comparable to the disk of the Milky Way.  The nuclear region
($R_{gal} < 200$ pc) and beyond $R_{gal}=800$ pc show an increased
value of $R_{13}$.  The material at small galactocentric radius may be
affected by the presence of low-luminosity AGN and a nuclear starburst
\citep{agn-classify} which results in an increase in the kinetic
temperature of the molecular gas.  Such an increase in the temperature
is thought to produce an increased value of $R_{13}$ near the nuclei
of many galaxies \citep{cotope-ratio} and is consistent with the
enhanced \co$(J=2\to1)$/\co$(J=1\to 0)$ ratio observed at
$R_{gal}\lesssim 100$ pc in the NUGA study.  At large radii, the
material may be disturbed by infalling material from the
counter-rotating atomic disk beyond $R_{gal}=1000$ pc because the
increase in $R_{13}$ is seen in gas where radial motions are observed
(\S\ref{rc}).  The change in $R_{13}$ is probably because the fraction
of gas in a diffuse, translucent phase is higher in this region than
in the remainder of the galaxy.  In both of these regions, the
standard \co-to-H$_{2}$ conversion factor is suspect, since it is
derived from gas where most of the molecular emission arises from
cold, bound, opaque GMCs.

Near $R_{gal}\sim 0$, the ratio $W_{\mathrm{HCN}}/W_{\mathrm{CO}}$ is
comparable to that seen in the nucleus of the Milky Way \citep{hcn-MW}
On average, the value of $W_{\mathrm{HCN}}/W_{\mathrm{CO}}$ over the
entire galaxy is comparable to the value seen in the high column
density regions of GMCs in the Solar neighborhood or the inner disk of
the Milky Way.  The relatively high values of this line ratio imply
that a substantially larger fraction of the molecular gas is found in
high density regions than is typical for the Solar neighborhood where
$W_{\mathrm{HCN}}/W_{\mathrm{CO}}\lesssim 0.01$ \citep{hcn-MW}.

\subsection{Fourier Analysis of the \co\ and \cotope\ Line Ratios}
The integrated intensity images of the \co\ and \cotope\ emission look
qualitatively different; the \cotope\ emission looks significantly
``clumpier.''  Since the \cotope\ emission should be a more faithful
tracer of molecular column density than \co\ emission, the clumps seen
in Figure \ref{cotopemap} suggest that there are clouds present in the
data.  However, the \cotope\ map also has lower dynamic range and the
effects of masking may produce the difference in structure.  In order
to determine whether the \cotope\ map is actually clumpier, we
measured the variation in line ratio as a function of spatial scale
using Fourier techniques.  Since bound clouds have a fixed spatial
scale ($\ell \sim 100$ pc in the Milky Way), emission arising from
them should have most of its power around this scale.  In contrast,
emission from diffuse clouds is not limited to a fixed scale and
should contribute at all spatial scales.  Since the line ratio
$R_{13}$ is different in diffuse vs. self-gravitating clouds, the
ratio of power in the \co\ map to that of the \cotope\ map should vary
as a function of spatial scale.

\begin{figure}
\begin{center}
\plotone{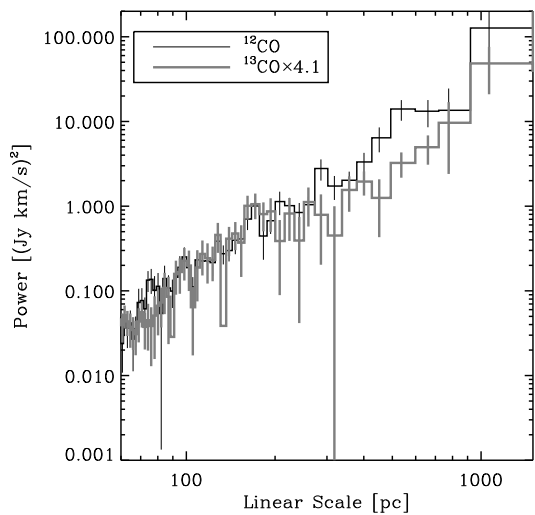}
\caption{\label{ftpow} Power spectrum of \co\ and \cotope\ images as a
function of spatial scale.  The figure compares the power spectrum of
the original \co\ data and the \cotope\ image after being scaled by
ratio of the fluxes at small spatial scales (4.1).  The distribution
of emission agrees well on the small scales but the line ratio at
large spatial scale scales is larger (6.5).  The difference in line
ratios as a function of scale implies that the clumpy structure in
Figure \ref{cotopemap} is real and the line ratios are comparable to
those seen in GMCs in the Milky Way.}
\end{center}
\end{figure}

We generated an integrated intensity map of the galaxy by shifting the
emission in the spectra to a common velocity in order to eliminate the
effects of galactic rotation.  We integrated over a window 106.25 km
s$^{-1}$ wide (25 channels) centered on the emission line.  This
generated integrated intensity maps with improved signal-to-noise
without introducing any erroneous spatial filtering from masking the
data.  The resulting maps contained $>80\%$ of the emission found in
the original maps with the missing emission in the nuclear region
where the linewidth exceeds the width of the window.  Next, we Fourier
transformed the integrated intensity maps and averaged the power in
elliptical annuli so that each annulus represents a range of scales in
the {\it deprojected} image.  We then subtracted the noise power from
the data using the transform of noise images generated from
emission-free regions of the data cube.  The resulting power as a
function of scale should represent only the power from line emission.
In Figure \ref{ftpow} we plot the results of the analysis, scaling the
\cotope\ image up by the ratio of the fluxes at small scales (4.1) to
facilitate comparison.  The power spectra have the same shape at large
scales ($\ell > 200$ pc) and small scales ($75 < \ell < 200$ pc) but
they have different ratios of powers in these two regimes: $6.5 \pm
0.3$ for $\ell > 200$ pc and $4.1 \pm 0.1$ for $70 < \ell < 200$ pc.
The value of $R_{13}$ at small scales is comparable to that seen in
Galactic GMCs, and emission on these scales is from structures that
may be the analogues of Galactic GMCs.  Compared to the \co\ map, the
\cotope\ map has less emission at large scales where only diffuse gas
should contribute.  Hence, \cotope\ emission from clouds will be less
confused by the presence of diffuse gas compared to \co\ emission.

The flux ratios derived in this analysis do not agree with the ratios
calculated for the galaxy as a whole (Figure \ref{linerat}) for two
reasons: (1) much of the emission with a large value of $R_{13}$ is
located in the center of the galaxy where the velocity range is
smaller than the linewidth so some high-$R_{13}$ emission is not included;
and (2) incomplete subtraction of the noise power will skew the ratios
to lower values.  This is because the noise power, relative to the
signal, is larger in the \cotope\ emission.  If there is incomplete
subtraction of the noise, it will contribute more to the power in the
\cotope\ emission than to the \co\ emission.

The observed difference in flux ratios as a function of scale persists
even when the dynamic range of the \co\ image is reduced to match that
of the \cotope\ image.  We conclude that the different morphologies in
Figures \ref{comap} and \ref{cotopemap} represent a real difference in
the structure of the emission from the galaxy and are not the result
of the masking process.  Furthermore, it appears that there is
significant contribution to the molecular emission from a diffuse,
translucent component of the molecular ISM. By analogy with molecular
emission in the Milky Way, it is likely that it has a conversion
factor $X_2 < 1$.  Despite the lower signal-to-noise, the \cotope\ map
should be a better tracer of the molecular emission from bound clouds
in M64.  We return to the question of diffuse gas in \S \ref{diffuse}.

\section{Decomposition of the Molecular Emission}
\label{decomp}
To answer the fundamental question of this study, namely `Are there
GMCs in M64?', we decompose the blended molecular emission into clouds
and then compare the properties of these clouds with Local Group GMCs.
For this purpose, we focus on the \cotope\ emission since this tracer
will be less contaminated by emission from diffuse molecular gas.
There are three principal methods for isolating structures in blended
molecular emission (1) partitioning of the cube by eye
\citep[e.g.,][]{ws90,ic342.wright}, (2) ``cleaning'' the data set by
iteratively fitting and subtracting three dimensional Gaussians
pioneered with the GAUSSCLUMPS algorithm by \citet{gaussclumps} (3)
partitioning of the emission by assigning it to the ``closest'' local
maximum first developed in the CLUMPFIND algorithm by
\citet{clumpfind}.  We adopt the latter approach for M64 since it does
not require known functional forms for the \cotope\ distribution of
the clouds and it works directly with the emission present in the
datacube.  The original CLUMPFIND algorithm was implemented for the
high signal-to-noise regime on single dish data.  We have redeveloped
the algorithm using the methods of \citet{clumpfind-mod} to make the
method more robust in the presence of noise.  In addition, we have
modified the distance metric in the algorithm to be applicable to
interferometer data which is oversampled in the spatial dimension.  A
detailed discussion of the algorithm appears in Appendix
\ref{algorithm}, including benchmarking information for the choice of
parameters. The algorithm can be applied to noisy, interferometric
data with minimal variation in the properties of the derived clouds as
a result of small changes in the free parameters of the algorithm.

Despite the modification of the algorithm to minimize the effects of
of noise, decomposition of the \cotope\ masked data used in Figure 3
was hampered by including low significance emission and noise.  To
improve the sensitivity of the datasets, we smoothed all three
datacubes to a channel width of 8.5 km~s$^{-1}$.  We regenerated a
mask for the \co\ data using the same method described in
\S\ref{linemaps}, namely we required $I>2\sigma_{rms}$ in a pixel that
is connected to $I>4\sigma_{rms}$ detections in two adjacent channels.
However, for the \cotope\ and HCN data, we required $I>2\sigma_{rms}$
in two adjacent channels where there was detectable CO emission.  The
threshold was chosen so that $<1$ false detection would be included in
the \cotope\ and HCN masks.  While this method discriminates against
low significance emission, it does a far better job of selecting only
real emission in the weaker tracers of molecular gas.

\begin{figure*}
\begin{center}
\plotone{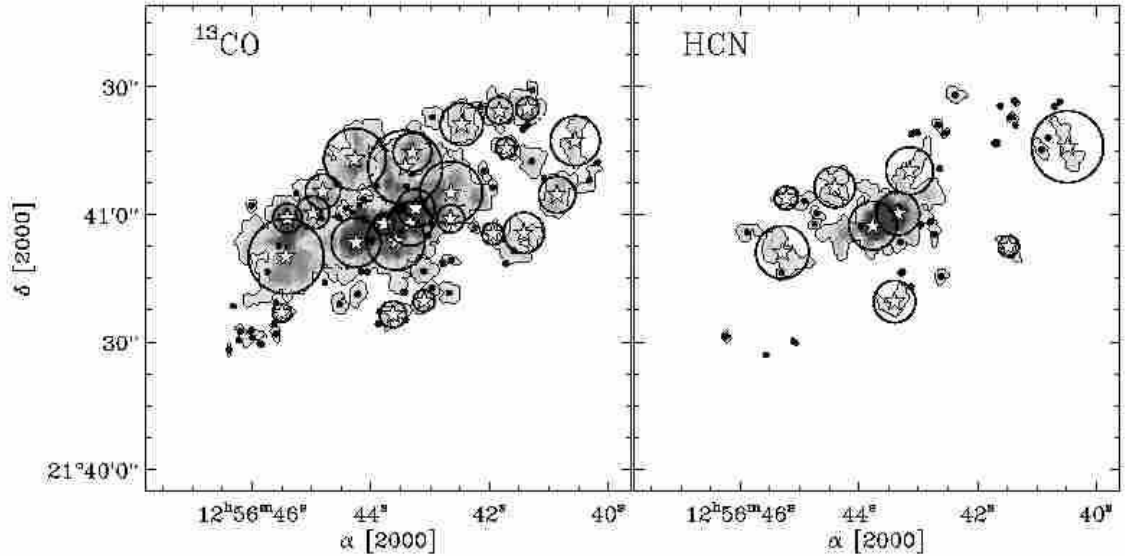}
\caption{\label{cloudmaps} The results of the decomposition algorithm
applied to the \cotope\ and HCN data sets.  The locations of the
clouds are overlaid on a grayscale integrated intensity maps of the
galaxy.  Locations of the clouds are indicated by stars (for clouds
that are resolved in both position and velocity) and dots (unresolved
in either position or velocity).  The algorithm identifies 87 clouds
in the \cotope\ data, of which 25 are resolved, and 42 clouds in the
HCN data, of which 9 are resolved.  The effective radii ($R_e$, see
Appendix \ref{cloudmeth}) of resolved clouds are indicated by the
circle drawn around them.  Several of the \cotope\ clouds are also
seen in the HCN emission with similar derived properties.}
\end{center}
\end{figure*}

We applied the decomposition algorithm to the molecular emission in
all three tracers.  We also applied the algorithm to the \co\ data at
the full velocity resolution because of the high signal-to-noise of
the original CO data.  In Figure \ref{cloudmaps}, we display the
masked, integrated intensity images for \cotope\ and HCN, including
the locations of the clouds into which the emission is decomposed.
The algorithm isolates 87 clouds in the \cotope\ emission and 42
clouds in the HCN.  Only a fraction (25 and 9, respectively) of these
clouds are resolved in both position and velocity space and these
clouds are indicated with stars.  The decomposition of the \co\ data
at full velocity resolution yields 44 clouds.  The algorithm
identifies several clouds that appear in all three tracers with
similar properties.

\subsection{Cloud Properties}
\label{cloudprops}

We measure the macroscopic properties of these clouds using
intensity-weighted moments of the emission distribution, following the
methods of \citet{srby87} and \citet{rpeb03} to facilitate comparison
with Local Group GMCs.  Using these methods, we determine the mean
position and line-of-sight velocity of each cloud.  In addition, we
measure three macroscopic properties for the clouds: the velocity FWHM
($\Delta V$), luminosity ($L$) and deconvolved effective radius
($R_e$).  We have corrected the derived values for these properties to
account for the fact that the data are clipped at
$T_{clip}=2\sigma_{rms}$; and therefore, the properties are not
measured over the same dynamic range as for Local Group clouds.  To
facilitate comparison with other data sets, we adopt the Gaussian
correction used in \citet{smcgmcs} and \citet{gmcs-galcen}.  The
derivation of the macroscopic cloud properties is given in detail in
Appendix \ref{cloudmeth}.  Table \ref{cloudlist} is a listing of the
cloud properties for the \cotope\ clouds.

\begin{deluxetable*}{cccccccc}
\tablecaption{\label{cloudlist}Properties of the resolved 
\cotope\ clouds}
\tabletypesize{\footnotesize}
\tablewidth{0pt}
\tablehead{
\colhead{Number} & \colhead{Position\tablenotemark{a}}
& \colhead{$V_{LSR}$} & \colhead{$R_e$} & \colhead{$\Delta V$} &
\colhead{$L_{\mathrm{13CO}}$} & \colhead{$P$\tablenotemark{b}} \\
& \colhead{($'',''$)} & [km s$^{-1}$] & 
\colhead{[pc]} & \colhead{[km s$^{-1}$]} 
& \colhead{[$10^4$ K km s$^{-1}$ pc$^{2}$]} & \\
}
\startdata
1 & (29,--7) & 266.1 & 176 & 28 & 87.4 & 4.7 \\
2 & (10,--4) & 292.8 & 115 & 48 & 79.7 & 5.1 \\
3 & (--4,13) & 473.5 & 174 & 68\tablenotemark{c} & 67.0 & 3.6 \\
4 & (--16,7) & 569.3 & 146 & 26 & 56.6 & 4.1 \\
5 & (--6,4) & 539.1 & 89 & 39 & 55.1 & 4.2 \\
6 & (--1,--4) & 401.6 & 136 & 73\tablenotemark{c} & 54.8 & 3.8 \\
7 & (--5,0) & 491.8 & 104 & 46 & 53.4 & 3.6 \\
8 & (10,15) & 399.0 & 144 & 38 & 42.0 & 4.0 \\
9 & (28,1) & 298.0 & 68 & 21 & 31.8 & 4.5 \\
10 & (2,0) & 345.1 & 50 & 35 & 23.4 & 3.5 \\
11 & (19,8) & 349.8 & 81 & 23 & 21.5 & 3.2 \\
12 & (22,3) & 318.7 & 78 & 19 & 16.3 & 3.4 \\
13 & (--45,7) & 550.6 & 86 & 18 & 15.8 & 3.3 \\
14 & (--19,24) & 506.4 & 99 & 25 & 13.4 & 2.2 \\
15 & (--37,27) & 544.0 & 53 & 18 & 13.4 & 3.8 \\
16 & (--16,1) & 541.8 & 64 & 28 & 12.9 & 3.1 \\
17 & (--36,--2) & 519.6 & 95 & 16 & 10.7 & 1.8 \\
18 & (--50,20) & 581.3 & 115 & 16 & 10.3 & 2.3 \\
19 & (--29,27) & 520.6 & 64 & 16 & 9.6 & 3.5 \\
20 & (--5,17) & 493.3 & 90 & 19 & 8.2 & 1.7 \\
21 & (--9,--18) & 404.6 & 50 & 10 & 7.2 & 2.5 \\
22 & (--31,18) & 546.5 & 48 & 20 & 7.0 & 2.2 \\
23 & (0,--21) & 380.3 & 62 & 23 & 5.4 & 2.4 \\
24 & (--27,--2) & 509.4 & 51 & 12 & 4.9 & 1.9 \\
25 & (30,--21) & 264.1 & 43 & 16 & 4.2 & 1.7 \\
\enddata
\tablenotetext{a}{Given in arcseconds relative to the center of the
galaxy at $\alpha_{2000}=~12^{\mbox{\scriptsize h}}\ 56^{\mbox{\scriptsize m}}\ 43\fs 6$ and
$\delta_{2000}=21\degr\ 40\arcmin\ 57\farcs 7$}
\tablenotetext{b}{The ratio of the peak antenna temperature to the
clipping level for the clouds, which determines the magnitude of the
Gaussian correction applied (see Appendix \ref{cloudmeth}).}
\tablenotetext{c}{This cloud is a blend of two peaks of emission which
cannot be separated objectively by the algorithm.}
\end{deluxetable*}

\subsection{Comparison with Milky Way GMCs}
\label{gmcs}
\begin{figure*}
\begin{center}
\plottwo{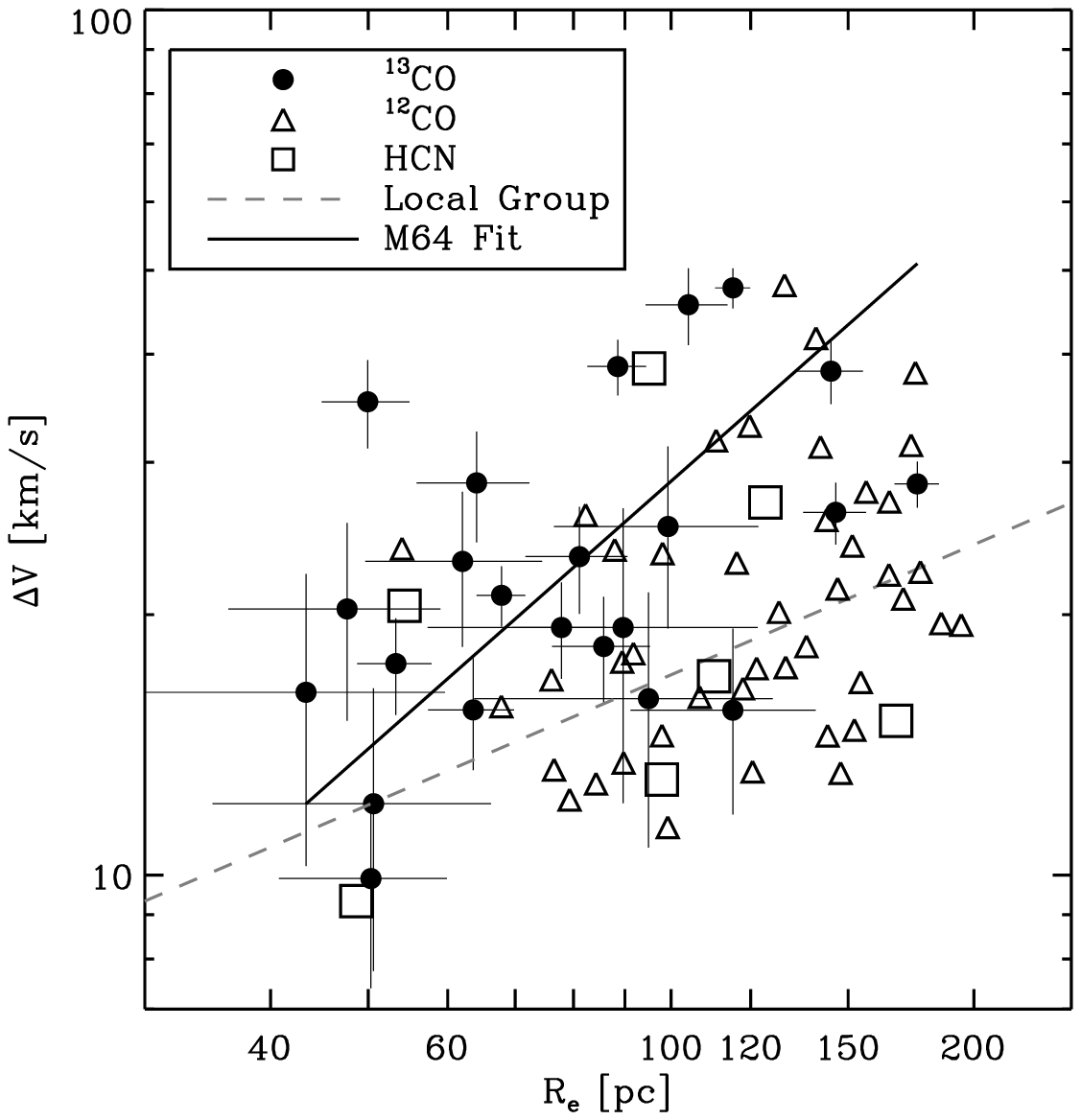}{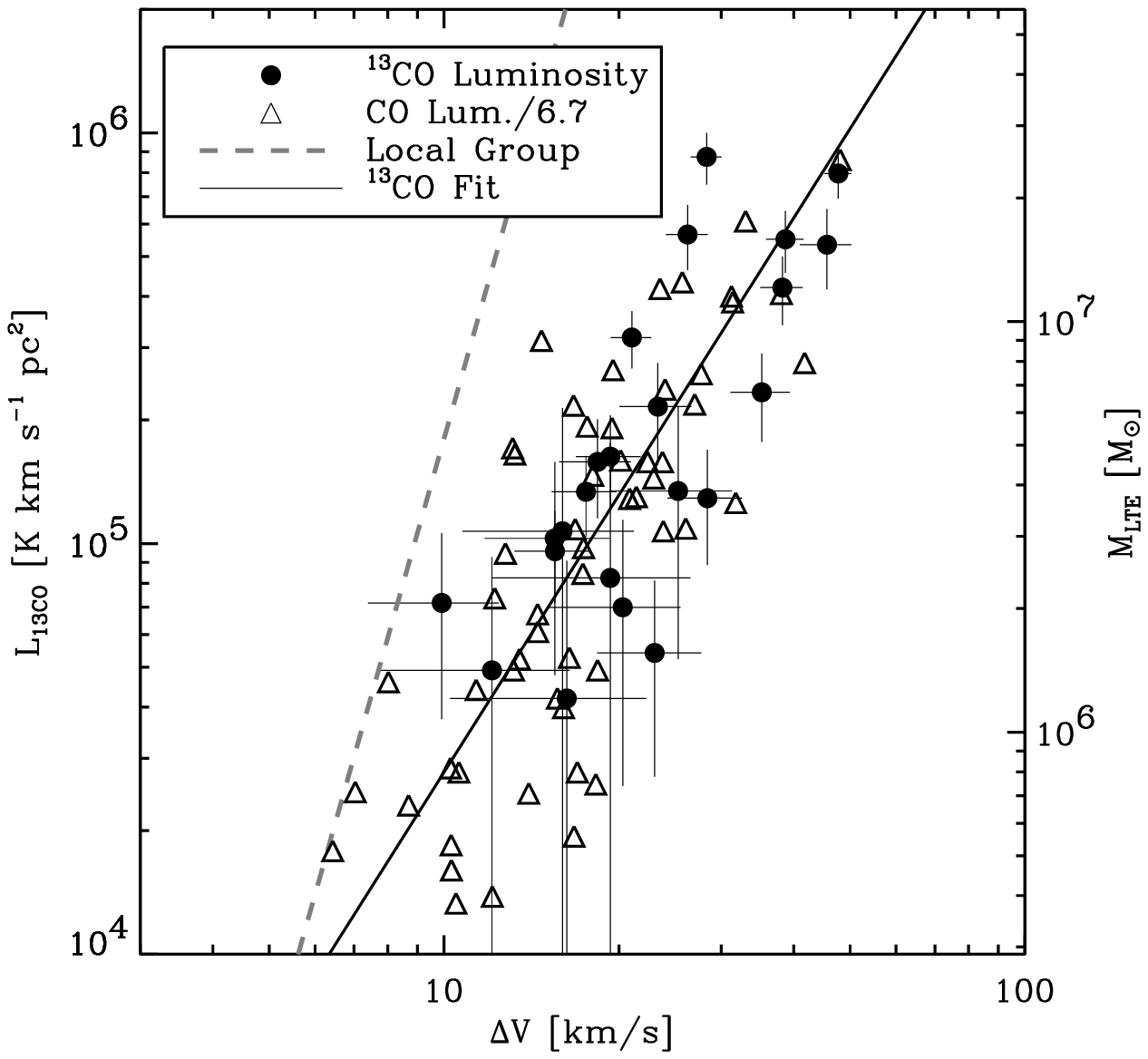}
\caption{\label{larson} Plot of the linewidth--size and
luminosity--linewidth relationship for clouds in M64.  The left-hand
panel plots the relationship between cloud sizes and linewidths from
the analysis of all three tracers.  The right-hand panel plots the
luminosity as a function of linewidth for the \co\ and \cotope\
clouds.  The luminosities of the \co\ clouds have been scaled down by
6.7 to compare them with the \cotope\ clouds.  The derived properties
of the clouds are different from the corresponding Local Group trends
plotted as dashed, gray lines.  The M64 clouds appear to have smaller
sizes or larger linewidths (or both) than clouds in the Local Group.
For reference, the right-hand axis of the luminosity-linewidth plot
indicates the luminous masses of the clouds assuming $X_2=1$ and
$R_{13}=6.7$.}
\end{center}
\end{figure*}

In the Local Group, the macroscopic properties of molecular clouds are
related by power law relationships, first noted by \citet{larson}.  In
particular, the study of \citet{srby87} found
\begin{eqnarray*}
\left(\frac{\Delta V}{\mbox{km s}^{-1}}\right)=1.7 
\left(\frac{R_e}{\mbox{pc}}\right)^{0.5}\mbox{ and }\\
\left(\frac{L_{\mathrm{12CO}}}{\mbox{K km s$^{-1}$ pc$^{-2}$}}\right) = 1.8
\left(\frac{\Delta V}{\mbox{km s}^{-1}}\right)^5.
\end{eqnarray*}
We use these values for comparison with the results of this study
since we have attempted to replicate their measurement methods and
systematic effects. To compare the derived \cotope\ luminosities to
the luminosities of S87, which are measured in $^{12}$CO, we scale
their relationship down by a factor of $R_{13}=6.7$, the global
\co/\cotope\ line ratio, in the inner disk of the Milky Way
\citep{mw-1213}. In Figure \ref{larson}, we plot the relationships
between these macroscopic properties for the clouds in M64.  We have
excluded clouds with linewidths larger than $\Delta V > 60\mbox{ km
s}^{-1}$.  These clouds contain multiple peaks and appear to be the
blend of two smaller clouds.  These clouds are found near the center
of the galaxy where the large velocity gradient makes decomposition of
the emission difficult.

Several features are immediately apparent from Figure \ref{larson}.
First, the properties of the clouds are significantly different from
the relationships derived by S87.  Second, the clouds have large
luminosities compared to the Local Group clouds.  In S87, very few
clouds are massive enough to have a \cotope\ luminosity larger than
$10^5\mbox{ K km s$^{-1}$ pc$^{2}$}$.  For reference, the mass implied
by the \co\ luminosity has also been plotted in Figure \ref{larson}
assuming $R_{13}=6.7$ and $X_2 =1$, which gives a \cotope-to-H$_2$
conversion factor equivalent to that derived using dust extinction by
\citet{lada-x13}.  To compare the Milky Way and M64 clouds in a more
quantitative fashion, we fit linear trends between $\log(R_e)$,
$\log(\Delta V)$ and $\log(L_{\mathrm{13CO}})$ for the \cotope\ clouds
using the method of \citet{bces}, which is appropriate for a
population with intrinsic scatter and measurements with errors in both
parameters.  We find that
\begin{eqnarray*}
\log \left(\Delta V\right) &=& 
(-0.6\pm 0.4)+(1.0\pm 0.3) \log \left(R_e\right) \mbox{ and }\\
\log \left(L_{\mathrm{13CO}}\right) &=& 
(2.2\pm 0.6)+(2.2\pm 0.4) \log \left(\Delta V\right).
\end{eqnarray*}
The index for the linewidth--size is marginally different from the
value for the Local Group ($1.0\pm 0.3$ vs.~0.5), and the
luminosity-linewidth is significantly different from the Local Group
trend ($2.2\pm 0.4$ vs.~5.0).  These fits indicate that the clouds
identified in the M64 data are not like molecular clouds seen in the
disk of the Milky Way.  However, the luminosity of these clouds
implies that most of the clouds are significantly more massive than
any clouds analyzed in the S87 study, and indeed \citet{wm97} argue
that no clouds more massive than $M>6\times 10^6~M_{\odot}$ will be
found in the Milky Way disk.  Thus, we are comparing the properties of
these clouds with an {\em extrapolation} of Local Group clouds to high
mass.  It is noteworthy, however, that the low luminosity clouds in
M64 have comparable luminosities, sizes and linewidths to the most
massive clouds in the Milky Way.  Thus, the M64 clouds may indicate a
change in the scalings among GMC properties that occurs at high mass
(or high average surface density).  We cannot determine whether the
apparent differences with respect to the S87 trends are an
environmental effect or are intrinsic to high mass molecular clouds.

While the analysis has focused on the \cotope\ clouds, Figure
\ref{larson} shows that the clouds generated for other tracers have
similar behaviors.  Scaling the luminosity of the CO clouds down by
6.7 to place them on a similar scale as the \cotope\ clouds produces a
luminosity--linewidth relationship statistically indistinguishable
from the relationship for the \cotope\ clouds.  In contrast, the sizes
of the \co\ clouds appear significantly larger than the \cotope\
clouds.  A two-sided KS test indicates that the distributions of the
\co\ and \cotope\ cloud radii are significantly different
($P_{KS}>0.9993$), while the distributions of the linewidths are
indistinguishable.  We suspect that the larger \co\ radius results
from the algorithm including the diffuse \co\ emission that surrounds
the clouds in the definition of the \co\ clouds.  To confirm this, we
re-cataloged the \co\ emission clipping the map at $5\sigma_{rms}$,
producing roughly the same dynamic range as the \cotope\ map.  The
high clipping level eliminates the low surface brightness, diffuse
\co\ emission.  Our inclusion of a Gaussian extrapolation to a common
clipping level facilitates direct comparison between the resulting
catalogs.  The clipped catalog contained 33 clouds; these clouds have
sizes comparable to the \cotope\ clouds, and a two-sided KS test found
the distributions of both the sizes and the linewidths to be
indistinguishable between the two tracers.  The presence of diffuse
emission can thus significantly affect the sizes of clouds traced by
\co\ emission alone.

A major concern with the decomposition is whether the structures that
the algorithm isolates really do represent discrete physical entities
or whether the ``clouds'' are really blends of several small clouds
that cannot be decomposed with the resolution of these observations.
We suggest that this is not the case and that we are cataloging
physical objects for several reasons.  First, we tested the hypothesis
that the clouds are blends of several GMCs like those found in the
Milky Way by analyzing simulated data sets.  For each data set, we
constructed a population of 50 molecular clouds with a total mass of
$2\times 10^{7}~M_{\odot}$ and individual masses drawn randomly from a
mass distribution with $dN/dM \propto M^{-1.6}$.  For each cloud in
the simulation, we calculated the radius and linewidth that would be
expected based on the study of S87.  We randomly distributed these
clouds in a data cube so that they would appear as a cloud with
$R_e=$150 pc and $\Delta V = 40$~km~s$^{-1}$, which is appropriate for
a typical massive cloud in our data set (Figure \ref{larson}).  We
then convolved the simulated cube to the resolution of the
observations and added noise to produce a peak signal-to-noise ratio
of 8.  We processed each simulated cube with the same masking and
decomposition algorithm used in the data analysis.  The algorithm
identifies substructure within the data cubes in 90\% of the trials,
leading us to conclude that the clouds are not likely to be blends of
a GMC population like that seen in the disk of the Milky Way.

The second piece of evidence that the decomposition products represent
real objects is that the linewidths of the observed clouds, while
large, are still significantly smaller than would be expected if the
emission were composed of several smaller clouds orbiting in the
galactic potential.  We estimate the linewidth due to shear as 
\[\Delta V_{shear}^2 = \frac{\sum_i I(x_i,y_i)\cdot 
(\bar{v}-v_{r})^2(x_i,y_i)} {\sum_i I(x_i,y_i)}+\Delta V^2(x_0,y_0),\]
where $I(x_i,y_i)$ is the integrated intensity at each position in
the cloud, $(\bar{v}-v_r)(x_i,y_i)$ is difference between the mean
velocity and the line-of-sight projection of the rotation
velocity at each position in the cloud, and $\Delta V^2(x_0,y_0)$ is
the linewidth of the emission distribution at the cloud center.  The
final term accounts for broadening due to beam-smearing and the
intrinsic gas velocity dispersion.  The measured linewidth for the
clouds ($\Delta V$ in Table \ref{cloudlist}) is smaller than $\Delta
V_{shear}$ by a factor of 2.3 on average and does not change
significantly with radius, even in the outer galaxy where the shear
decreases.

\begin{figure}
\begin{center}
\plotone{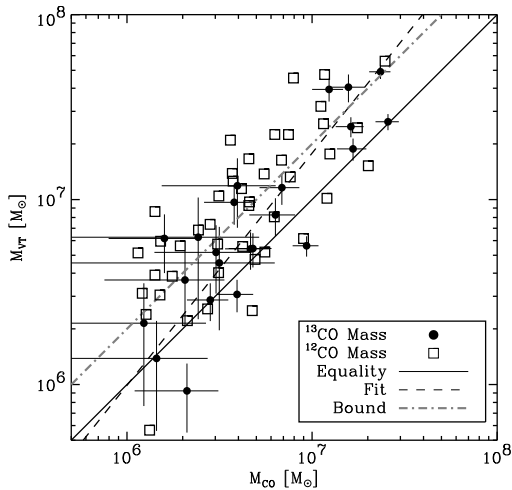}
\caption{\label{vtplot} Relationship between virial mass estimates and
luminous mass estimates of the molecular gas for both \cotope\ and
\co\ clouds.  The luminous mass estimates assume $X_2=1$ and $R_{13}=6.7$.
The good correspondence between mass estimates for over an order of
magnitude in mass implies that the clouds isolated by the algorithm
are discrete physical entities, illustrated by a linear fit between
the $\log(M_{\mathrm{VT}})$ and $\log(M_{\mathrm{CO}})$.  The locus of
marginally bound objects for $X_2=1$ and $R_{13}=6.7$ is shown as the
dot-dashed gray line.}
\end{center}
\end{figure}

Estimates of the cloud masses using the virial theorem agree well with
estimates using the \cotope\ emission, giving a third piece of
evidence that we are identifying clouds.  In particular,
if we adopt the virial mass estimate used by S87,
\begin{equation}
\left(\frac{M_{\mathrm{VT}}}{M_{\odot}}\right) = 187
\left(\frac{\Delta V}{\mbox{km s}^{-1}}\right)^2 
\left(\frac{R_e}{\mbox{pc}}\right),
\end{equation}
we find good correspondence between the virial mass estimates and
luminous mass estimates as indicated in Figure \ref{vtplot}.  A linear
fit between the log of the two mass estimates gives
\[\log(M_{\mathrm{VT}}) = (-1.5\pm 1.2) + (1.2\pm 0.2) 
\log(M_{\mathrm{CO}}),\]
consistent with a direct correspondence between the mass estimates
over an order of magnitude.  This good agreement again suggests that
the clouds are discrete objects, since the application of the virial
theorem to unbound groups of molecular clouds gives virial masses
significantly in excess of the mass estimated from the CO luminosity
\citep[e.g.,][]{rand-gma-m51,rand-gma-m100,allen-m31}.  We have also
used four alternate methods to measure the virial parameters
\citep{alt-virial} and find no significant differences between the
various methods.

If the clouds represent self-gravitating objects, we can use the
relationship between the luminous mass estimates and dynamical mass
estimates to derive conversion factors for the tracers.  Assuming the
clouds are virialized, we perform a least-squares fit between
$\log(M_{\mathrm{VT}})$ and $\log(L_{\mathrm{CO}})$, fixing the slope
at unity to determine the scaling between these two tracers.  Using
these fits, we derive CO-to-H$_2$ conversion factors:
\[ N(\mbox{H}_2) = (1.8\pm 0.2) \times 10^{21}\mbox{ cm}^{-2} 
\cdot W_{\mathrm{13CO}}\]
\[ N(\mbox{H}_2) = (3.2 \pm 0.2) \times 10^{20}\mbox{ cm}^{-2}
\cdot W_{\mathrm{12CO}}\mbox{ or } X_2=(1.6\pm 0.1),\] where the
surface brightness $W$ is expressed in K~km~s$^{-1}$.  If we assume
the clouds are marginally bound instead of being virialized, we derive
conversion factors that are a 2 times lower than these values.  Given
that the derived conversion factors bracket the measured values for
both \co\ \citep{dht01,sm96} and \cotope\ \citep{lada-x13}, it appears
that dynamical estimates of the mass agree well with masses estimated
from CO luminosity.  Significant external pressure or the presence of
magnetic fields will reduce the value of $M_{\mathrm{VT}}$, closing
the small gap between the luminous ($X_2=1$, $R_{13}=6.7$) and virial mass
estimates \citep{mckee-vt}.  In light of this, we adopt $X_2=1$ for
the \co\ observations and the conversion factor of \citet{lada-x13}
for the \cotope.  Under the assumption of virialization, we also
derive
\[\left(\frac{M}{M_{\odot}}\right) = (90\pm 10) 
\left(\frac{L_{\mathrm{HCN}}}{\mbox{K km s$^{-1}$
pc$^{-2}$}}\right),\] implying HCN is brighter relative to CO in these
clouds than is seen in Solar neighborhood GMCs \citep{hcn-MW}.
However, HCN/CO ratio is similar to gas in the nuclear disk of the
Milky Way where the mean density is higher.

\begin{figure}
\begin{center}
\plotone{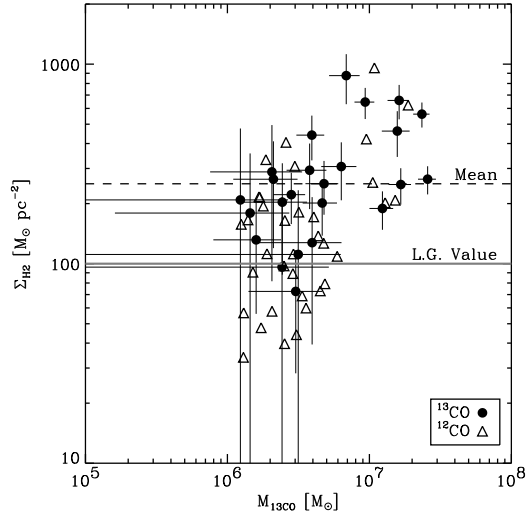}
\caption{\label{sdplot} The scaling of molecular surface density
($\Sigma_{\mathrm{H2}}$) with luminous mass $M_{\mathrm{13CO}}$.  The
molecular clouds in M64 have significantly higher surface densities
than do clouds in the Milky Way.  Moreover, the surface density is not
constant with mass, as is necessary if the clouds are self-gravitating
but follow a different size-linewidth relationship from clouds in the
Local Group.}
\end{center}
\end{figure}

With mass measurements of the molecular clouds, we can convert these
estimates to surface densities of the clouds.  We plot the surface
density of molecular gas in the clouds as a function of cloud mass in
Figure \ref{sdplot}.  On average, the surface densities of the clouds
are larger than are seen in the Milky Way with a mean value of
$(250\pm 15)~M_{\odot}~\mbox{pc}^{-2}$, significantly higher than the
value of $100~M_{\odot}~\mbox{pc}^{-2}$ seen in the Milky Way
\citep{psp3} and M33 \citep{rpeb03}.  In addition, the surface density
is {\it not} constant with mass, which must be true if the clouds in
the galaxies are self-gravitating but do not follow the same linewidth--size
relationship as is observed in the Local Group.  We find that
\[\Sigma_{\mathrm{H2}}\propto M^{0.7\pm 0.2}\]
meaning that high mass molecular clouds are systematically smaller and
denser than would be expected from the extrapolation of trends seen in
the Local Group.  Moreover, an increased surface density of molecular
gas also implies that the internal pressures of the molecular clouds
are larger than are typical in the Local Group.  We estimate the
internal pressures of these clouds:
\begin{eqnarray}
\frac{P_{int}}{k} &\sim&\langle \rho \rangle \sigma_v^2\nonumber\\
&=& 212 \mbox{
cm$^{-3}$ K} \left(\frac{M}{M_{\odot}}\right)\left(
\frac{R_e}{\mbox{pc}} \right)^{-3} \left(\frac{\Delta V}{\mbox{km
s}^{-1}}\right)^2.
\label{press_eq}
\end{eqnarray} 
We find a mean internal pressure of $10^6$~K~cm$^{-3}$, an order of
magnitude larger than the value inferred for Milky Way GMCs
\citep{psp3}.  However, an important quantity for comparison with the
internal pressure is the mean external pressure exerted on the cloud
by the ISM of M64.  \citet{pressure1} formulated an expression for the
midplane hydrostatic pressure of the disks of galaxies:
\begin{eqnarray}
\frac{P_{ext}}{k} &= &272 \mbox{ cm$^{-3}$ K} 
\left(\frac{\Sigma_{gas}}{M_{\odot}\mbox{ pc}^{-2}}\right) 
\left(\frac{\Sigma_{*}}{M_{\odot}\mbox{ pc}^{-2}}\right)^{0.5}\nonumber\\
&&\cdot\left(\frac{v_g}{\mbox{km s}^{-1}}\right)
 \left(\frac{h_*}{\mbox{pc}}\right)^{-0.5}
\label{pext}
\end{eqnarray}
where $v_g$ is the 1-dimensional velocity dispersion of the gas, $h_*$
is the stellar scale height, and $\Sigma_*$ is the stellar surface
density.  We assume a gas velocity dispersion of $v_g=8$ km s$^{-1}$,
corresponding to the linewidth of forbidden line emission in
\citet{m64_stellar}.  We derive an upper limit for the pressure by
assuming and $h_*$ of 300 pc, typical of disk galaxies
\citep{pressure1}.  We measure $\Sigma_*$ derived from the 2MASS Large
Galaxy Atlas image of M64 \citep{2mass-lga} assuming $M_K/L_K=0.5$
\citep{m2l-bell}.  Finally, we take $\Sigma_g =
\Sigma_{\mathrm{H2}}+\Sigma_{\mathrm{HI}}$.  We calculate
$\Sigma_{\mathrm{H2}}$ from the \cotope\ surface density profile and
use $\Sigma_{\mathrm{HI}}=10~M_{\odot}\mbox{ pc}^{-2}$ as measured in
\citet{counterrot_m64}.  With these data, we compare the internal
pressures of molecular clouds to the external pressure at the position
of the cloud and find that the clouds are overpressured relative to
the external medium by {\it at least} a factor of 2.

If we refine this estimate, the difference between $P_{int}$ and
$P_{ext}$ grows larger.  Of particular note, this estimate has
included the contribution of the gas in self-gravitating molecular
clouds in the contribution to the external pressure.  These clouds
likely do not contribute to the hydrostatic support of the ISM.  Since
this gas dominates the local column density, the actual external
pressure is likely significantly smaller than the above estimate.  If
we use the estimate that 25\% of the molecular gas is found in the
diffuse component (\S\ref{diffuse}) and assume that self-gravitating
clouds do not contribute to the hydrostatic support of the ISM, then
the clouds have internal pressures higher than the local ISM by a
factor of 12, similar to the margin by which Solar neighborhood GMCs
are overpressured with respect to the local ISM \citep{psp3}.  Recent
star formation activity will decrease the mass-to-light ratio and
therefore the mass estimate of the stellar disk, also reducing the
external pressure.  The central concentration of mass in GMCs tends to
increase the internal pressures.  Finally, the stellar scale height is
likely larger than the 300 pc typical of galactic disks because of the
presence of a significant bulge in the galaxy across the inner $\sim
300$ pc \citep{disk_fits}.

Because the clouds are significantly overpressured with respect to the
ambient medium, we take this as a final piece of evidence that the
clouds are self-gravitating and that the luminosity traces molecular
mass with conversion factors comparable to Milky Way values.  Adding
this to the agreement between the luminous and dynamical mass
estimates and the small linewidths of the clouds, there is good
evidence that molecular clouds identified by this method are
gravitationally bound.  The masses of many clouds are significantly
larger than are seen in the disk of the Milky Way, where galactic
tides set an upper limit to the mass of a GMC \citep{tides}.  There is
significant shear in the inner disk of M64 which may disrupt the
clouds so we measured the tidal acceleration across each of the GMCs.
We find that the clouds in Table \ref{cloudlist} are marginally stable
against tides using the methods of \citet{tides}.  All clouds at
$R_{gal} > 400$~pc are very stable against tidal disruption.  We
conclude that these clouds are the analogues of the GMCs seen in the
Milky Way, but have significantly different macroscopic properties.

\subsection{Robustness of the Results}

Several assumptions and choices have been incorporated in the
conclusion that the molecular clouds observed in M64 are
self-gravitating structures, analogous to high mass Giant Molecular
Clouds.  However, the principal results of our study do not change
even if we relax these assumptions.

The largest influence on our cloud properties is the choice of the
decomposition algorithm.  The algorithm we developed has been designed
to be robust against small variations in its parameters.  Although we
justify the choice of free parameters in Appendix \ref{algorithm}, we
have re-cataloged the \cotope\ emission while varying contouring
levels and values of $T_{uniq}$ (the free parameters of the
algorithm).  We found all changes in the cloud properties were small
and well within the uncertainties.

Because the clouds are comparable to the beam size, there is
significant concern that the objects we catalog are the blends of
several smaller clouds.  We have argued in \S\ref{gmcs} that this is
not the case, based on simulations, pressure arguments, dynamical mass
estimates and cloud linewidths.  Several of our results rely upon
accurate decomposition of the emission.  However, the most important
of our results remain even if the clouds are blends of smaller
structures.  Namely, the surface (and volume) densities of the small
clouds must be even higher than measured for the M64 clouds, implying
they are even more over-pressured with respect to the ambient ISM; the
estimate of the internal pressure (Equation \ref{press_eq}) becomes a
lower limit.  Moreover, much of the following discussion
(\S\ref{disc}) is independent of how the clouds are decomposed.  We
also emphasize that the molecular structures cannot consist of Milky
Way GMCs overlapping along the line of sight since Milky Way GMCs
clouds are not dense enough to exist as discrete structures in the
high-pressure molecular disk of M64.

The derived properties have been corrected for the finite
signal-to-noise ratio in the dataset using a Gaussian extrapolation
(see Appendix \ref{cloudmeth}).  We have tested the validity of the
Gaussian correction to the cloud properties by fitting
three-dimensional Gaussians to the results of the decomposition.  We
find that the extrapolated values agree well with the fits.  If we
repeat the analyses of \S\ref{gmcs} without the corrections to their
properties, we find (1) the indices of the linewidth--size and
luminosity--linewidth relationships are still significantly different
from the Local Group values, (2) the luminous and virial mass
estimates still agree well, (3) the mean surface density of the clouds
is $(450\pm 20)~M_{\odot}\mbox{ pc}^{-2}$, even higher than our
adopted value and (4) the clouds are still significantly overpressured
with respect to the ISM of M64.

Finally, the distance to M64 is somewhat uncertain with a recent
estimate of $D=7.5$ Mpc \citep{sbf-tonry}.  Adopting a larger distance
will change the values of the luminosity (as $D^2$) and the sizes (as
$D$) but leave the linewidths unaffected.  The indices of the scalings
between the macroscopic properties will be invariant and remain
different from the Local Group. The clouds will appear to have higher
luminous masses relative to their virial masses
($M_{\mathrm{VT}}/M_{\mathrm{CO}}\propto D^{-1}$) and will still
appear bound, though the derived conversion factors will be smaller.
The surface densities of the molecular clouds do not vary with the
adopted distance and will be significantly higher than Local Group
values.  While changing the adopted distance will alter the measured
cloud properties, the main conclusions are robust: the most massive
clouds in M64 appear to be self-gravitating structures with
macroscopic properties significantly different from an extrapolation
from Local Group GMCs.

\section{Discussion}
\label{disc}
In this section, we relate the results of this study to molecular gas
on galactic scales.  We estimate the contribution of diffuse molecular
gas to emission from M64 and examine the dust content and star-forming
properties of the molecular clouds.  Clarifying the properties of
these high mass molecular clouds helps to link our understanding of
star formation on small scales in the Milky Way to starbursts and
ultraluminous infrared galaxies (ULIRGs).

\subsection{The Fraction of Diffuse Gas in M64}
\label{diffuse}
Given the large \co\ to \cotope\ line ratio ($R_{13}$) observed in
some portions of M64, we suspect that a significant contribution of
the emission arises from diffuse, translucent molecular gas.  The
resolved GMCs in Table \ref{cloudlist} comprise $\sim$50\% of the
\cotope\ emission from the galaxy.  We cannot distinguish how much of
the remaining emission arises from GMCs below our completeness limit
or whether the emission arises from diffuse molecular gas.

Since emission from diffuse gas has a substantially different value of
$R_{13}$ from that found in the GMCs, we can use our observations to
establish the ratio of \co\ emission coming from diffuse gas vs. bound
objects ($F\equiv W_{diff}/W_{GMC}$):
\[F=\frac{R_{diff}}{R_{GMC}}\frac{R_{avg}-R_{GMC}}{R_{diff}-R_{avg}}\]
where $R_{diff}$ is the line ratio in diffuse gas, $R_{GMC}$ is the
line ratio in GMCs and $R_{avg}$ is the average of the line ratio over
a given region \citep{mw-1213,m33-1213}.  Within the 25 clouds
identified in the \cotope\ emission, the line ratio $R_{13}=3.7$,
which is typical of the peaks of emission from GMCs in our own galaxy
\citep{mw-1213}.  This suggests that the physical conditions averaged
over these GMCs are comparable to the highest column density regions
of GMCs in our own Galaxy.  We assume that $R_{diff}=10\to 20$, which
is appropriate for high-latitude clouds \citep{highlat,highlat2} and
small clouds in the ISM \citep{tiny-clouds}. Then we can use the data
in Figure \ref{linerat} to estimate the fraction of diffuse emission
in the different regions of the galaxy.  From 200 pc $< R_{gal} <$ 800
pc, $R_{avg}\sim 6$ which implies, under these assumptions, that
40$\to$50\% of the \co\ emission comes from GMCs and the balance comes
from translucent molecular gas, comparable to the fraction seen in the
Milky Way. A significant fraction (20\%) of the \co\ emission comes
from the inner 200 pc of the galaxy, around the AGN.  Here, $R_{13}=8$
which implies that between 65\% and 85\% of the \co\ emission from the
nucleus arises from diffuse molecular material which is equal to the
fraction over the entire galaxy.  In contrast, between 25\% and 60\%
of the \cotope\ emission comes from diffuse gas.  Assuming the
\cotope\ emission linearly traces the molecular mass of the galaxy, we
assume $R_{diff}=20$ and estimate that 25\% of the molecular mass
resides in diffuse structures and the remainder is in bound clouds.

The results of the Fourier analysis of the emission suggest a similar
trend.  In the \cotope\ emission, the power on small ($\ell < 200$~pc)
scales comprises 20\% of the total power, but for \co, this decreases
to 10\%.  The Fourier analysis does not completely separate diffuse
from bound emission since the diffuse component of the molecular ISM
contributes power on small scales as well.

Line ratios indicate that a substantial portion of the molecular
emission arises from diffuse gas, where the standard CO-to-H$_2$
conversion may be suspect.  The \cotope-to-H$_2$ conversion factor is
more appropriate for estimating the mass of the galaxy.  Using
$X_2=1$, we find that $M_{\mathrm{CO}}=(3.2\pm 0.4)\times
10^{8}~M_{\odot}$.  Using the \cotope-to-H$_2$ conversion factor of
\citet{lada-x13}, we find $M_{\mathrm{13CO}}=(2.7\pm 0.1)\times
10^{8}~M_{\odot}$, roughly 30\% of the Milky Way value
\citep{sigma_dame}.  The two mass estimates for the molecular gas
agree within the errors in the flux measurements, and the mass
measurement of the gas appears robust.  Using \cotope\ emission to
measure molecular mass will likely provide a more robust measurement
of molecular mass in star-forming molecular clouds rather than the
diffuse gas which is presumably inert with respect to star-formation.

\subsection{Dust and Extinction in M64}
Both CO isotopomers confirm a large mass of molecular gas in the inner
kiloparsec of M64; and with the implied large column density of
molecular material, there should be accompanying dust.
\citet{tdust-spirals} use ISO and SCUBA data to measure the long
wavelength ($\lambda > 100\mu\mbox{m}$) infrared and submillimeter
emission from M64 and find an emission distribution consistent with
two dust components at $T_{kin}=14$~K and 28~K respectively.  The
total dust mass implied by their observations is $2.7\times
10^6~M_{\odot}$ which implies a gas-to-dust ratio of $\sim 100$ by
mass, very close to Galactic \citep{dust2gas}.  If the molecular gas
and the corresponding dust were distributed uniformly, the extinction
through the disk should be $\sim A_V=7$.  Stellar extinction
measurements imply $A_V\sim 1.5$ \citep[NUGA, ][]{m64-extinction}
implying that the dust (and gas) must have a filling fraction smaller
than unity.  The interferometer data show $\langle T_{A,max} \rangle =
1.7$~K for CO but it is unlikely that the gas is significantly colder
than $T_{kin}=10$~K or warmer than $T_{dust}=14~$K.  If the difference
is because of beam dilution, we estimate that the typical covering
fraction of the gas in GMCs is $15\%\lesssim f \lesssim 45\%$.

\subsection{Star Formation in M64}
Large surface densities of molecular gas are linked to high star
formation rates \citep{k98,wb02}.  In this section, we synthesize the
results of the observations in several wavebands to determine if the
star forming properties of the molecular gas in M64 are peculiar.

Using the FIR fluxes from the IRAS Bright Galaxy Sample
\citep{iras-bgs}, and the scalings of \citet{total_fir}, we derive an
FIR flux of $1.9\times 10^{-9}$ erg cm$^{-2}$ Hz$^{-1}$ from
$S_{60}=30.2$ Jy and $S_{100}=78.7$ Jy.  Scaling this with the
relationship of \citet{k98} yields a total SFR of 0.17 $M_{\odot}$
yr$^{-1}$ for $R_{gal} < 1$~kpc or 0.05
$M_{\odot}$~yr$^{-1}$~kpc$^{-2}$, nearly an order of magnitude larger
than in the Solar neighborhood \citep{mw-sfr}.

We synthesized a radio continuum spectral energy distribution from
several studies in the literature
\citep{nvss,turner_ho,m64_rc1,m64_rc2} and derive
\[S(\nu)=(0.130\pm 0.005) \mathrm{\ Jy\ }\left(\frac{\nu}
{1.4\mbox{ GHz}}\right)^{-0.61\pm 0.03}.\] The radio continuum
emission is confined to the molecular gas disk \citep{nvss,turner_ho}
and is well within the primary beam in all the interferometric
studies.  The resulting SFR is $0.15\pm 0.04~M_{\odot}$ yr$^{-1}$
using the formula of \citet{condon_araa}.  Comparing this to the study
of \citet{murgia} indicates that M64 is quite similar to other
galaxies with similar surface densities of molecular gas.

H$\alpha$ is one of the most commonly used star formation tracers in
the galaxy, though using it requires correcting for extinction, which
can be large.  B94 use a wide field image and estimate the H$\alpha$
flux as $4.8\times 10^{-12}
\exp(\langle\tau_{\mathrm{H}\alpha}\rangle)$ erg s$^{-1}$ cm$^{-2}$,
where $\langle\tau_{\mathrm{H}\alpha}\rangle$ is the typical optical
depth for H$\alpha$ emission, taken by \citet{counterrot_m64} to be
1.3.  The corresponding star formation rate is 0.28 $M_{\odot}$
yr$^{-1}$ with the scalings of \citet{k98}.  The Pa$\alpha$ image
presented in \S\ref{recomb} has a total flux of $9.3 \times
10^{-13}\exp(\langle\tau_{\mathrm{Pa}\alpha}\rangle)$ erg s$^{-1}$
cm$^{-2}$.  For a comparable degree of extinction as assumed in
\citet{counterrot_m64}, $\langle\tau_{\mathrm{Pa}\alpha}\rangle=0.23$
with the extinction law of \citet{cardelli}.  This corresponds to a
total star formation rate of 0.15 $M_{\odot}$ yr$^{-1}$ with the
scaling of \citet{k98}. This is not unexpected since the image covers
a smaller area than the H$\alpha$ image. The recombination line
emission appears to accurately trace the star formation in the galaxy.
In order for the line-of-sight extinction to the \ion{H}{2} regions to
be only $A_V=1.7$ while having the gas imply a mean extinction of
$A_V=7$ again requires significant clumping of the molecular gas or
else essentially all the H$\alpha$ emission would be extinguished by
dust absorption.

The synthesis of observations suggest a star formation rate of $\sim$0.2
$M_{\odot}$ yr$^{-1}$, or 0.06 $M_{\odot}$ yr$^{-1}$ kpc$^{-2}$, over
the molecular, star-forming disk.  This agrees well with the high
surface density regions of other molecule rich galaxies
\citep{wb02,murgia}.  Given the large density of molecular material
present, the star formation activity is relatively quiescent, with a
molecular gas depletion time of $\tau \gtrsim 1.4$ Gyr.  Thus, the
star formation efficiency of the molecular clouds is similar to
that of the Milky Way \citep{ms88}.

\subsection{The GMCs as Star Forming Structures}

\begin{figure*}
\begin{center}
\plotone{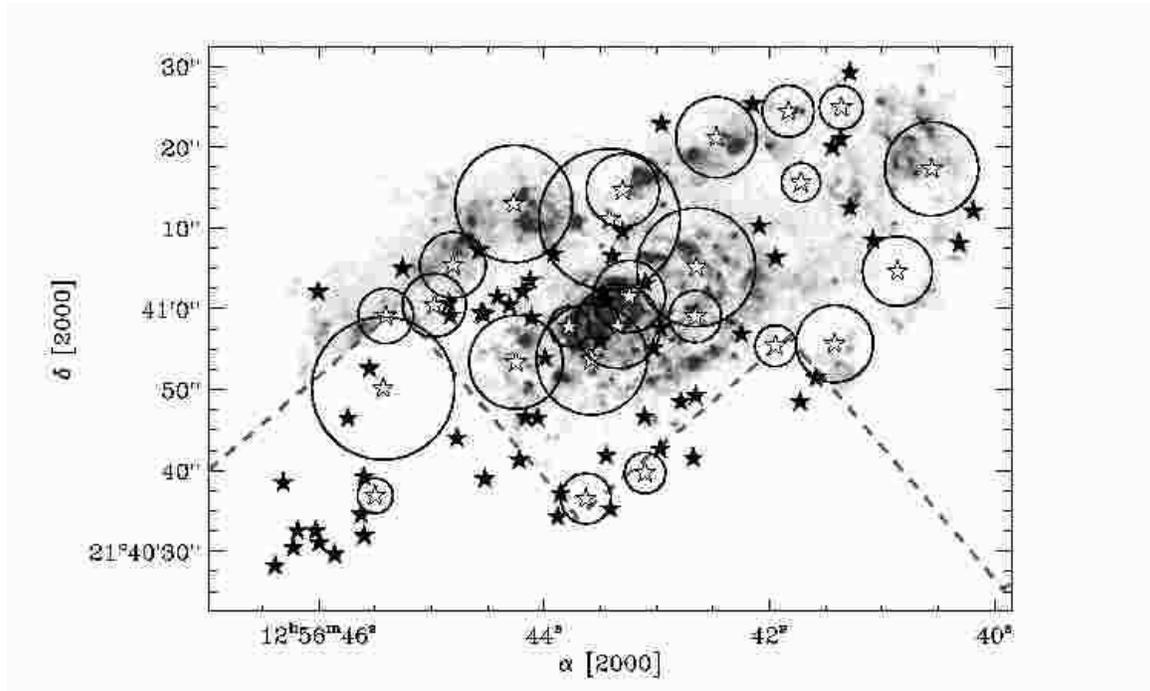}
\caption{\label{halpha} -- Locations of \cotope\ molecular clouds
overlaid on H$\alpha$ grayscale image.  Resolved clouds are indicated
with white stars and circles with of radius $R_e$ and unresolved
clouds indicated with black stars.  The border of the WFPC2 image is
shown as a gray dashed line. Across the molecular disk, there is a
correlation between GMCs and recombination emission, presumably from
star formation.}
\end{center}
\end{figure*}

In \S \ref{gmcs}, we argued that there are self-gravitating clouds of
molecular gas in M64, though their properties are substantially
different from the GMCs found in the disk of the Milky Way.  These
massive clouds also appear to be the star forming structures in M64.
In Figure \ref{halpha}, we indicate the positions of molecular clouds
from our catalog on the continuum subtracted HST image discussed in
\S\ref{recomb}.  Across the molecular disk, there is a strong
correlation between GMCs and recombination emission from star
formation.  To quantify this, we convolved the H$\alpha$ image to the
resolution of the \cotope\ data and compared the brightness on a pixel
by pixel basis.  The correlation coefficient is $0.74\pm 0.02$ and
rank correlation coefficient is $0.50\pm 0.03$.  Performing a similar
analysis with the Pa$\alpha$ image yields a correlation coefficient of
$0.70\pm 0.02$ and $0.65\pm 0.02$.  Star formation is clearly
correlated with the gas in the Giant Molecular Clouds, but measuring
the star formation rate within individual GMCs will require high
resolution FIR observations.  Unfortunately, the angular resolution of
Spitzer observations will not be sufficient to correlate the FIR flux
distribution with the location of GMCs.

Because the GMCs in M64 differ substantially from those of the Milky
Way, they may have significantly different star formation
efficiencies.  However, the global star formation efficiency is not
markedly enhanced.  Studies throughout the Milky Way
\citep{ms88,lck91,sch02} show that that the star formation efficiency
of clouds is constant with respect to cloud mass.  On a large scale,
it does not appear that the M64 clouds are peculiar in their star
formation efficiency.  This is not startling since even in ULIRGs
there is evidence that the star formation efficiency is constant
\citep{gs04}.  It is puzzling why such a large variation in the
internal properties of GMCs results in a strikingly constant star
formation efficiency.


\subsection{Comparison to other GMCs in High Density Environments}
\label{centercomp}

\begin{figure}
\begin{center}
\plotone{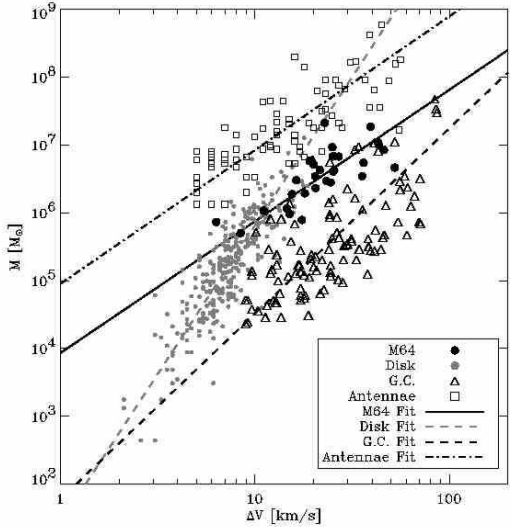}
\caption{\label{compprop} The mass--linewidth relationship for four
populations of GMCs.  The GMCs from M64 are compared with those found
in the Galactic center \citep{gmcs-galcen} the Antennae Galaxy
\citep{gma-wilson} and in the Galactic disk \citep[][S87]{srby87}. All
studies in high density systems have a similar mass--linewidth
relationship $M\propto\Delta V^{2}$, which is dramatically different
from the Milky Way relationship. The three high density cloud
populations have significantly different offsets in the relationship.}
\end{center}
\end{figure}

There are several published studies searching for molecular clouds in
high density environments.  In particular, we compare the results of
M64 to a CO study of the Antennae Galaxy (Arp 244) by
\citet{gma-wilson} and CO observations of the Galactic center by
\citet{gmcs-galcen}.  Although the studies use different decomposition
methods, it interesting to compare the results between the studies.
We plot the mass--linewidth relationship for the studies under
consideration in Figure \ref{compprop}.  We choose to compare the
mass--linewidth relationship since it avoids accurately measuring
deconvolved radii of clouds across several studies.  We include the
study of S87 for comparison with these three studies in high density
environments.  The masses derived from CO luminosity have been
rescaled to $X_2=1$.

Although all the studies use different methods for cloud
decomposition, we directly compare their derived clouds to the M64
sample.  \citet{gma-wilson} decomposed \co\ emission from the Antennae
galaxies using the original CLUMPFIND algorithm, isolating massive
molecular clouds.  Performing a linear fit between $\log(\Delta V)$
and $\log(M_{\mathrm{CO}})$ for their data yields an index of $2.0 \pm
0.1$, consistent with the M64 results and markedly different from the
Milky Way values.  The offset in luminosity is nearly an order of
magnitude larger than the M64 data.  \citet{gmcs-galcen} decomposed
the CO emission from the galactic center using the methods of S87.
The index of their relationship is $(2.7\pm 0.2)$ and the offset is
roughly an order of magnitude lower than the M64 clouds.  In all three
cases, the clouds follow a different mass-linewidth relationship than
do Milky Way disk clouds.

\begin{figure}
\begin{center}
\plotone{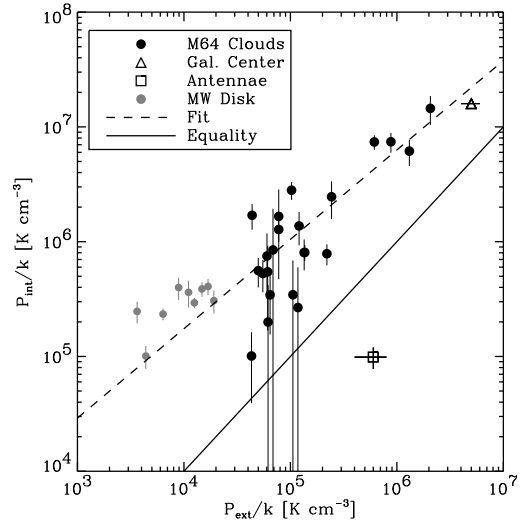}
\caption{\label{sdpress} A comparison of internal and external
pressures for Giant Molecular Clouds in a variety of systems.  Points
represent pressure estimates in clouds for M64 (present work), Milky
Way (S87), the Antennae galaxy \citep{gma-wilson}, and the Galactic
center \citep{gmcs-galcen}.  Internal pressures are estimated from
cloud properties.  External pressures are quoted values from X-ray
studies (Antennae, Galactic center) or estimated from the mass
contributions of galactic components (Milky Way, M64).  For most
clouds, the internal pressure scales with the external pressure and is
consistently above the derived value. }
\end{center}
\end{figure}

For 8 resolved clouds in the Antenna study, the virial masses agreed
with the luminous mass estimates (for $X_2=1.5$) and the average
surface density of the clouds is 100~$M_{\odot}\mbox{ pc}^{-2}$,
comparable to the clouds of S87.  On the opposite extreme, the mean
surface density of the molecular clouds in the Galactic center is
500~$M_{\odot}\mbox{ pc}^{-2}$, higher than M64 clouds by a factor of
2.  These differences are particularly interesting since they suggest
that the high surface densities observed in M64 are not intrinsic to
high mass GMCs but likely depend on environment.  One environmental
factor which may determine the properties of GMCs is the external
hydrostatic pressure.  GMCs cannot exist as discrete entities for much
longer than a sound crossing-time unless they are self-gravitating and
gravity is balanced by the internal pressure of the cloud.  In
environments that have high external pressures relative to the Milky
Way, the internal pressures of clouds should be proportionally larger
as well or else the pressure fluctuations in the diffuse ISM will
destroy the clouds.  To test the influence of environment on GMC
properties, we plot the internal pressure of GMCs as a function of the
external hydrostatic pressure for several galaxies (Figure
\ref{sdpress}).  For M64 and the disk of the Milky Way, we estimate
the pressure using methods described in \S\ref{decomp}.  For M64, we
assume that 1/4 of the molecular mass is diffuse and therefore
contributes to the external pressure with $X_2=0.25$.  We have
included a point representing the mean surface density for GMCs
estimated from the Galactic center data \citep{gmcs-galcen} at an
external pressure of $P_{ext}/k=5\times 10^6\mbox{ K cm}^{-3}$
\citep{pressure-galcen}.  We plot a point for the mean surface density
of the sample of \citet{gma-wilson} using their estimate of the
pressure, $P_{int}/k=6\times 10^5\mbox{ K cm}^{-3}$.  Finally, we
include the clouds of S87 and derive pressure values using the formula
of \citet{pressure1}, the stellar density of \citet{db98} and the gas
profile of \citet{sigma_dame}.  We have averaged the internal
pressures of the clouds in bins of $\Delta P_{ext}/k = 2\times
10^3\mbox{ K cm}^{-3}$ to highlight the significant scaling of
internal pressure with external pressure {\it within} the S87 sample.
This scaling is present using both luminous and dynamical mass
estimates of the GMC in the Milky Way and implies a systematic
variation of GMC column density with Galactic environment.  There
appears to be some offset between these populations, but this may be
due to different methods for evaluating the pressure.  The point
representing the mean value for the Antennae galaxies is notably
discrepant from the trend.  This may be because the objects cataloged
by \citet{gma-wilson} are physically distinct from those found in the
other samples, perhaps because of the large spatial resolution of
their study ($\theta_{beam}\sim 380$~pc).  Omitting the point for the
Antennae, we find a significant scaling with pressure: $P_{int}
\propto P_{ext}^{0.75\pm 0.05}$.  This scaling suggests that GMCs in
molecule-rich, high-pressure environments are characteristically
denser than those found in the Solar neighborhood.  Since GMCs in M64
also show scaling of surface density with molecular mass, it is
unclear as to whether the external pressure sets the surface densities
of the cloud or whether high pressure environments form high mass
clouds and the scaling with surface density is set by the cloud mass.
\begin{deluxetable*}{ccc}
\tablecaption{Summary of the GMC properties\label{sumtable}}
\tablewidth{0pt}
\tablehead{& \colhead{Milky Way} & \colhead{M64}}
\startdata
Maximum Mass [$M_{\odot}$] & $6\times 10^{6}$ & $2\times 10^7$\\
Size--Linewidth Scaling & $\Delta V \propto R_e^{0.5}$ & 
 $\Delta V \propto R_e^{1.0}$\\
Luminosity--Linewidth Scaling & $L\propto \Delta V^{5.0}$ & 
$L\propto \Delta V^{2.2}$ \\
Mass--Surface Density Scaling &  $\Sigma_{\mathrm{H2}}\propto M^{0.0}$
& $\Sigma_{\mathrm{H2}}\propto M^{0.7}$ \\
$\langle \Sigma_{\mathrm{H2}} \rangle$ [$M_{\odot}~\mbox{pc}^{-2}$] & 100 &
250\\
$\langle P_{int} \rangle$ [K cm$^{-3}$] & $10^5$ & $10^6$ \\
SFR [$~M_{\odot}$ yr$^{-1}$ kpc$^{-2}$] & 0.005 &  0.06 \\
Mol. Gas Depletion Time [Gyr] & 0.5 & 1.4\\
\enddata
\end{deluxetable*}
\section{Summary and Conclusions}

We have presented BIMA and FCRAO observations of the molecule rich
galaxy M64 (NGC 4826) in the $(J=1\to 0)$ transition of three tracers:
\co, \cotope\ and HCN.  The spatial resolution of the observations
projects to 75 pc at the distance of M64, making it feasible to
marginally resolve large molecular clouds in the galaxy.  We decompose
this emission into clouds that are high mass analogues of Giant
Molecular Clouds in the Local Group.  The properties of these clouds
are significantly different from those seen locally and are summarized
in Table \ref{sumtable}. We report the
following conclusions:

1. The line ratios $W_{\mathrm{CO}}/W_{\mathrm{13CO}}\equiv R_{13}$ and
$W_{\mathrm{HCN}}/W_{\mathrm{CO}}$ vary over the face of the galaxy.
The large values of $R_{13}$ seen in the nuclear regions of the galaxy
($R_{gal}<200$ pc) and beyond $R_{gal}=800$~pc imply substantial
emission from diffuse molecular gas. Between these two regions,
$R_{13}=6$, similar to the Milky Way disk. The ratio
$W_{\mathrm{HCN}}/W_{\mathrm{CO}}$ declines monotonically from 0.09 in
the middle of the galaxy to 0.012 at $R_{gal}=800$~pc.

2. A Fourier analysis of the emission distributions shows
that the value of $R_{13}=4.1$ on small scales ($\ell < 200$ pc) and
$R_{13}=6.5$ on large scales.  This implies that the clumpiness seen in the
\cotope\ image of the galaxy represents real cloud structures with
values of $R_{13}$ comparable to Milky Way GMCs.

3. We decomposed the \cotope\ emission into clouds and calculated
their properties.  We find several clouds with $M>10^7~M_{\odot}$
larger than any cloud seen in the disk of the Milky Way.  The clouds
show $L\propto \Delta V^{2.2\pm 0.4}$ which is a much smaller index
than is seen for clouds in the Milky Way disk where $L\propto \Delta
V^{5}$.  Similarly, we find $\Delta V \propto R_e^{1.0\pm 0.3}$ compared
to the Milky Way clouds where $\Delta V\propto R_e^{0.5}$.  Despite
these differences, virial estimates of their masses agree with those
derived from the \cotope\ emission for the Galactic \cotope\-to-H$_2$
conversion factor of \citet{lada-x13}.  The CO molecular clouds imply a
\co\-to-H$_2$ conversion factor of $(3.2\pm 0.2)\times 10^{20} \mbox{
cm}^{-2} \mbox{K~km~s}^{-1}$.

4.  High mass GMCs in M64 are significantly smaller and denser than
expected from extrapolating Local Group trends.  The mean surface
density of the M64 clouds is 250~$M_{\odot}\mbox{ pc}^{-2}$ and the
surface density of the clouds varies with cloud mass:
$\Sigma_{\mathrm{H2}}\propto M^{0.7\pm 0.2}$, compared to
$\Sigma_{\mathrm{H2}}\propto$~const. in the Local Group. Clouds with
masses comparable to those seen in the Milky Way also have comparable
linewidths and radii, suggesting that the clouds may be similar for
$M\sim 10^6~M_{\odot}$.

5.  Diffuse gas (i.e. not bound in GMCs) contributes roughly 2/3 of
the \co\ emission seen from the galaxy, but it only accounts for 1/4
the molecular mass in the galaxy.  Nonetheless, empirically derived
conversion factors for \co\ and \cotope\ emission give a similar value
of molecular mass for the entire galaxy: $2.7 \times 10^8~M_{\odot}$
(for $D=4.1$~Mpc) for $X_2=1$

6. Recombination line emission, FIR and radiocontinuum measurements of
the star formation rate M64 yield $\Sigma_{SFR}=0.06~M_{\odot}$
yr$^{-1}$ kpc$^{-2}$ or 0.2~$M_{\odot}$ yr$^{-1}$ for the central
molecular disk.  The molecular gas depletion time is $\sim$1.4 Gyr,
comparable to similar galaxies.

7. The GMCs found in M64 are star forming structures, judging from
their correlation with H$\alpha$ and Pa$\alpha$ emission in the
galaxy.  Despite the implied large values of their internal densities
and pressures, the star formation efficiency is not dramatically
enhanced.  The GMCs in M64 share many qualities with GMCs found in the
center of the Milky Way.

8.  The GMCs in M64 share many properties with populations of clouds
found in the inner Milky Way and the Antennae galaxy.  GMCs in all
three galaxies have a well-defined mass--linewidth relationship with
an index $\sim 2$, markedly different from the clouds in the inner
disk of the Milky Way.  We the internal pressure of GMCs scales with
the external pressure of the ISM: $P_{int}\propto P_{ext}^{0.75\pm 0.05}$.

\acknowledgements 

We thank Mark Heyer for invaluable assistance in conducting
observations using the FCRAO 14 m.  We thank David Meier for making a
copy of this thesis work on M64 available to us.  Fruitful discussions
with Adam Leroy and Alberto Bolatto greatly improved the methods used
in this paper.  We thank Tam Helfer for providing the BIMA SONG $UV$
data for M64.  The comments of an anonymous referee greatly improved
the presentation of the material.  The referee also suggested using
multiple ways of measuring virial masses.  ER's work was supported, in
part, by a NASA Graduate Student Research Program fellowship.  This
work would not have been possible without the extensive use of the
NASA Extragalactic Database, the NASA Abstract Data Service and the
HST data archive.  This work is partially supported by NSF grant
0228963 to the Radio Astronomy Laboratory at UC Berkeley.

\appendix
\section{The Modified CLUMPFIND Algorithm}
\label{algorithm}
\begin{figure}
\begin{center}
\epsscale{0.5}
\plotone{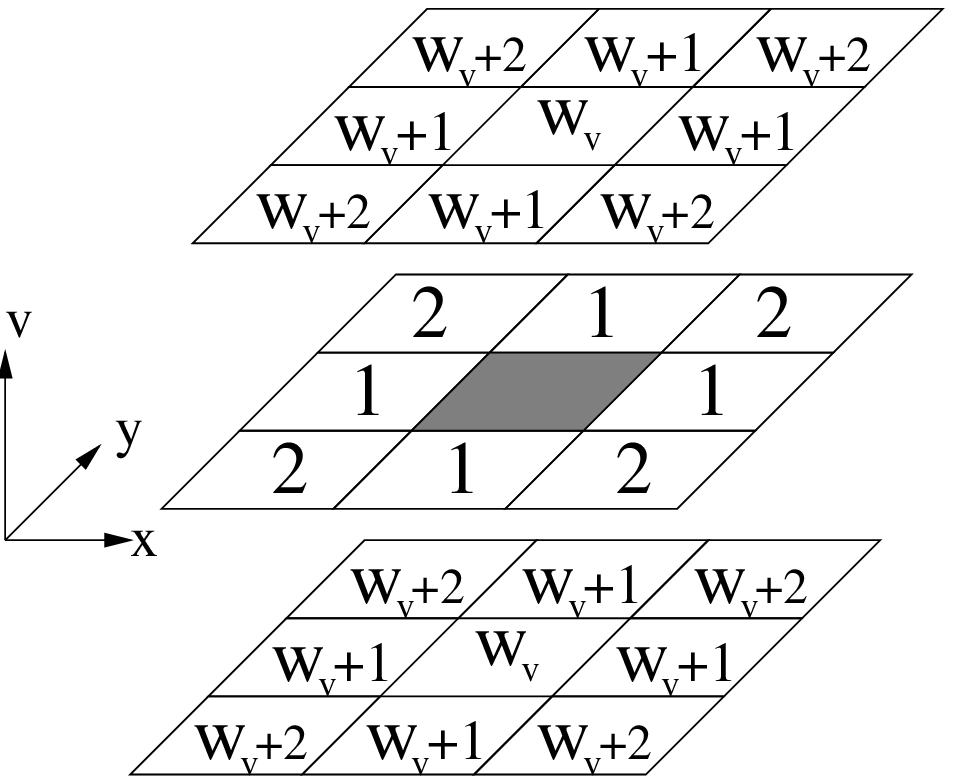}
\caption{\label{distances} Distances from the central shaded pixels to
adjacent pixels in the data cube.  The number $w_v$ accounts for the
oversampling in the spatial direction by making different velocity
planes correspondingly ``farther'' from the central pixel.  Within a
given plane, only the four pixels immediately adjacent to the central
pixel are considered neighbors and, consequently, one distance unit
away. This definition results in a ``taxicab metric'' in a given plane.}
\end{center}
\epsscale{1.0}
\end{figure}

In this paper, we use a modified CLUMPFIND \citep[][WGB94]{clumpfind}
algorithm since we seek maximum flexibility in identifying the
geometry of the molecular emission.  The dataset we are analyzing in
this paper is significantly different from that of WGB94, who used
single dish data that were spatially Nyquist sampled.  In contrast,
the interferometer data has several pixels across a beam and is often
at lower signal-to-noise than the data examined by WGB94.  To optimize
the CLUMPFIND method for our data, we adopt two significant changes.
First, we determine which local maxima to include in the analysis
before dividing the emission into clouds corresponding to those peaks.
This allows a more careful analysis in the relatively low
signal-to-noise domain.  The other change we adopt is to change the
path length metric so that path segments in velocity are a factor of
$w_v\equiv \theta_{FWHM} / 2\Delta x$ longer than path segments in
velocity.  $w_v$ is the number of spatial pixels per beam half-width,
roughly 4 in these data.  This change accounts for the oversampling in
the spatial direction of our data.  A graphic representation of the
distance metric adopted in the datacube is shown in Figure
\ref{distances}.

\begin{figure}
\begin{center}
\plottwo{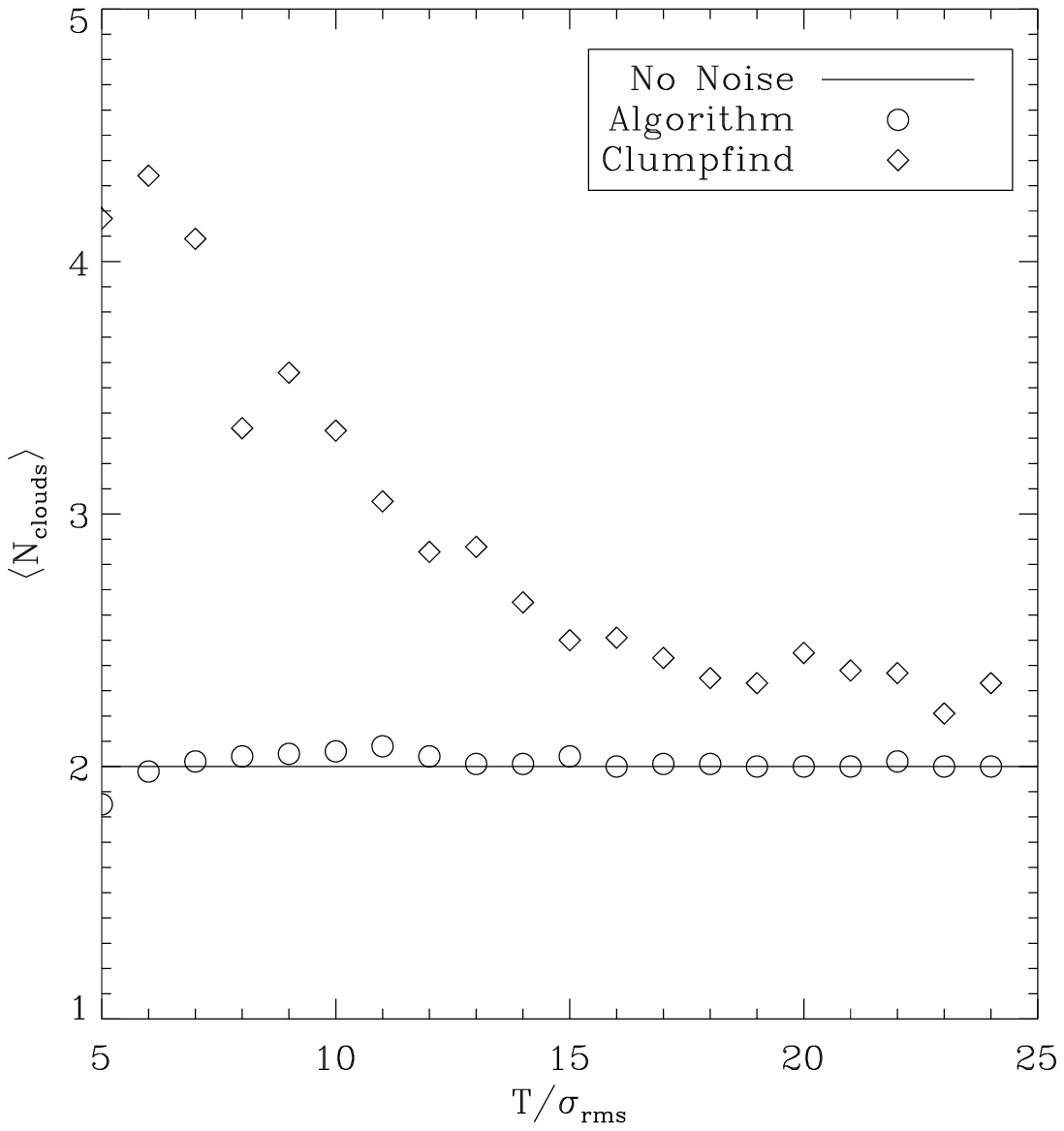}{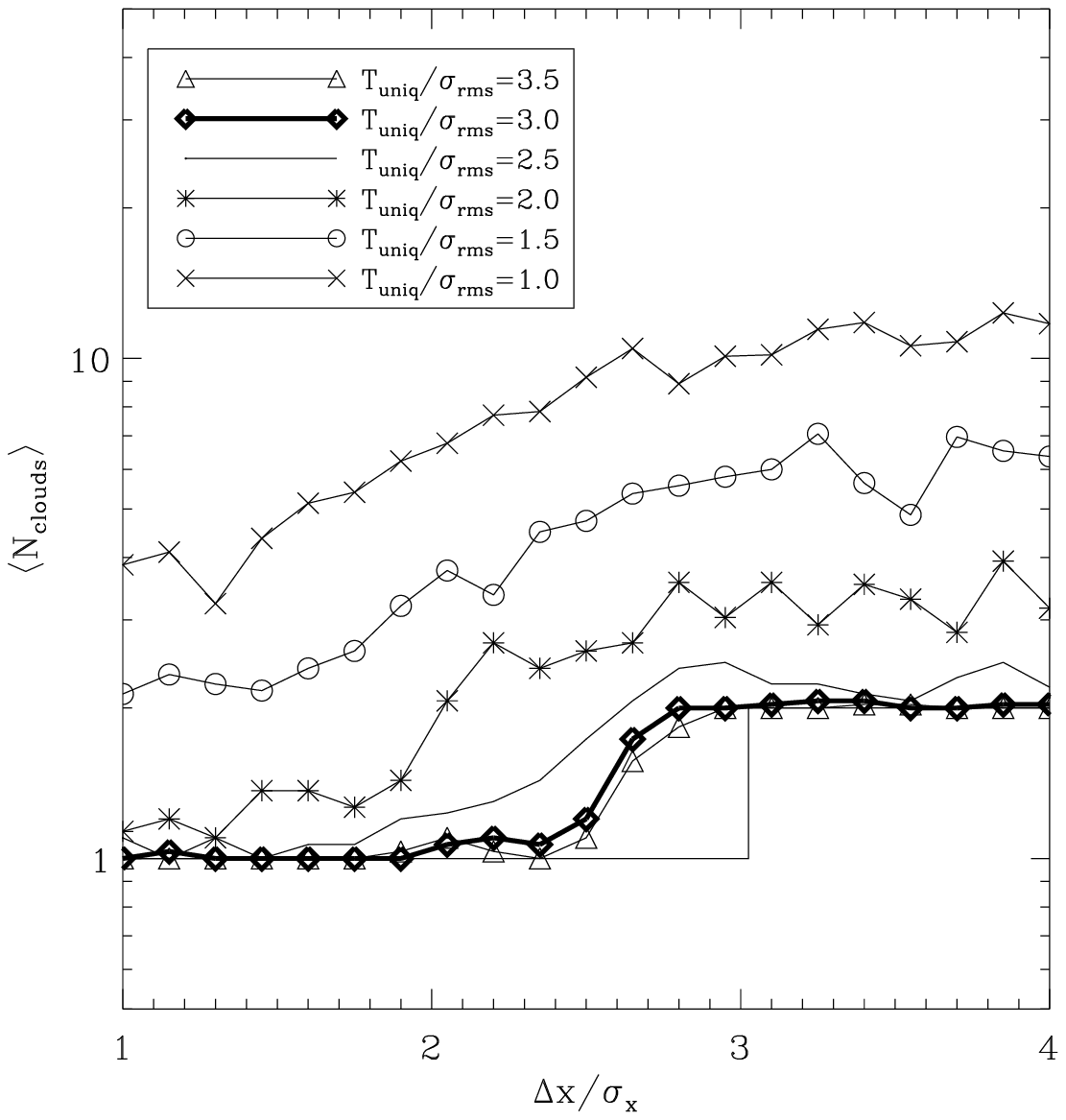}
\caption{\label{tuniq} Benchmarking results for the modified CLUMPFIND
algorithm.  In the left-hand panel we plot the number of clouds the
algorithm that result from decomposing a trial data set as a function
of signal to noise.  The modified algorithm produces good results even
down to a peak signal-to-noise value of 5.  In contrast, the CLUMPFIND
algorithm displays large sensitivity to noise in the low
signal-to-noise regime.  In the right-hand panel, we plot the number
of clouds that result from decomposition as a function of cloud
separation for $T_{peak}/\sigma_{rms}=8$.  The trials are conducted
using a range of values for the algorithm parameter $\Delta
T_{uniq}/\sigma_{rms}$, which determines the algorithm's sensitivity
to noise.  The trials show that $\Delta T_{uniq}/\sigma_{rms}=3$ is
reproduces the model well while not requiring extremely high
signal-to-noise in the data and that for this choice, the algorithm
separates the two model clouds for $\Delta x/\sigma_{x}\gtrsim 3$.}
\end{center}
\end{figure}

Accurate selection of significant local maxima is the key to
implementing our algorithm.  In WGB94, significant local maxima are
determined by contouring the data in units of $\Delta T /
\sigma_{rms}=2$, descending through the contour levels, and marking
any new, isolated island at a given contour level as a new maximum.
However, a noise fluctuation can cause an artificial local maximum and
a false clump and the choice $\Delta T / \sigma_{rms}=2$ is a
compromise between discriminating against noise spikes (large
contouring interval) and sensitivity to emission structure (small
contour level).  To optimize the algorithm, we separate these two
criteria: adopting a small contouring level to follow the emission
structure well while keeping track of the levels at which isolated
regions merge in order to later determine whether certain ``clouds''
represent noise spikes.  We adopt a similar method to (and heavily
inspired by) the work of \citet{clumpfind-mod}.  In their method, two
local maximum are considered distinct if the saddle point connecting
them is at least $\Delta T_{uniq} / \sigma_{rms}$ below both maxima.
\citet{clumpfind-mod} use $\Delta T_{uniq}/\sigma_{rms} = 4$ but we
find a value of 3.0 is appropriate for our data (see Figure
\ref{tuniq}).  In our method, we contour the data in units of
$0.5\sigma_{rms}$, and record isolated islands that appear in each
contour level as well as the level at which pairs of maxima merge.  If
a maximum merges with another, higher maximum and the saddle point
between the maxima is less than $\Delta T_{uniq}/\sigma_{rms}$, it is
considered to be the same cloud and the smaller local maximum is
removed from the analysis.  This selects against noise spikes since
they appear as isolated maxima in a single resolution element of the
datacube whereas cloud structures appear in multiple elements.  Once
all connected local maxima are tested for the presence of noise, the
remaining local maxima are used to partition the data set.

The data set is partitioned by stepping downward from the maximum
value of $T_A$ in units of $0.5\sigma_{rms}$.  Pixels added at each level
are assigned to the local maximum that has the shortest path through
the datacube connecting the pixel to that maximum.  Paths are
constrained to run through pixels which contain emission, and the
length of a path between two pixels is defined as the minimum length
over all possible paths that lie within the bounds of the emission.
Steps in velocity are considered to be a factor of $w_v =
\theta_{FWHM} / 2\Delta x$ longer than steps in position space.  If
pixel is connected to two maxima by paths of equal length, then the
pixel is assigned to the closest pixel using a Euclidean distance
metric:
\begin{equation}
d(\mathbf{x}_1,\mathbf{x}_2)=\sqrt{(x_1-x_2)^2+(y_1-y_2)^2+
	w_v^2(v_1-v_2)^2}.
\end{equation}
Here, $w_v$ is the same factor that accounts for oversampling in the
spatial direction.  If the pixels have equivalent path and Euclidean
distances to two separate maxima, the pixel is assigned randomly to
one.  

To evaluate the optimal parameters in running the algorithm, we
performed Monte Carlo simulations on controlled simulations.  The
simulated data consisted of two clouds with Gaussian profiles in
position and velocity space.  Interferometer noise was added to some
trials to mimic the observation conditions.  Ideally, the algorithm
will consistently separate the blended Gaussians into (only) two
clouds.  By varying the signal-to-noise, Gaussian separation and
algorithm parameters, we explored the limitations of the algorithm and
the ideal parameters for decompositions.  In Figure \ref{tuniq}, we
display the results of two such experiments.  First, we fixed the
cloud separation at $3\sigma_{x}$ and varied the peak signal-to-noise
to determine the limitations imposed by noise on the decomposition.
For $\Delta T_{uniq}/\sigma_{rms}=3$, we find that our modified
CLUMPFIND algorithm consistently recovers two clouds while the
original CLUMPFIND finds a significantly larger number owing to the
presence of noise.  The modified algorithm breaks down when the
signal-to-noise ratio is less than 5.  In the second panel of Figure
\ref{tuniq}, we compare the number of clouds resulting from the
decomposition as a function of cloud separation and $\Delta
T_{uniq}/\sigma_{rms}$.  This experiment explores the ability of the
algorithm to separate blended clouds and motivates our selection of
$\Delta T_{uniq}/\sigma_{rms}=3$.  We choose a signal-to-noise ratio
of 8 for the trials.  Selecting $\Delta T_{uniq}/\sigma_{rms}<3$
results in noise spikes causing a spurious decomposition of the
clouds.  Selecting $\Delta T_{uniq}/\sigma_{rms}>3$ does not
significantly change the results of the decomposition.   The algorithm
successfully separates clouds with a separation between them of
$\Delta x/\sigma_{x}\gtrsim 3$.  

We have also checked the influence on choice of contour levels and
find that a contouring the data in levels smaller 0.5 $\sigma_{rms}$
does not change the outcome of the decomposition.  We use
0.5$\sigma_{rms}$ in our analysis of the data.  In addition, we also
check the ratio of fluxes in the resulting clouds for trial clouds
with unequal intensities; and we find that the algorithm partitions
the clouds reasonably.

\section{Derivation of Cloud Properties}
\label{cloudmeth}
Using intensity-weighted moments, we have calculated the macroscopic
properties of the clouds in M64.  The position of the cloud is taken
as the centroid position, weighted by the antenna temperature in each
pixel $T_i$, e.g.,  $x_0=(\sum_i T_i x_i)/(\sum_i T_i)$.  The
linewidth and radius of the cloud are calculated by taking the second
moment of the emission in velocity and position respectively:
\begin{eqnarray}
\sigma_r^2 &=& \frac{1}{\sum_i T_i}\sum_i T_i 
\left[\left(x-x_0\right)^2+\left(y-y_0\right)^2 \right]\\
\sigma_v^2 &=& \frac{1}{\sum_i T_i}\sum_i T_i \left(v-v_0\right)^2\label{sigv}
\end{eqnarray}
To transform the moments into cloud radii and linewidths, we first
corrected the derived moments for the effects of clipping, since the
lowest contour level is at $T_A=2\sigma_{rms}$ rather than zero.  As
such, the derived moments underestimate the true $\sigma_r^2$ and
$\sigma_v^2$ of the cloud.  In order to compensate, we assumed the
clouds have a Gaussian profile in position-position-velocity space.
Since the clouds are only marginally resolved at best, the assumption
of a Gaussian profile seems to be reasonable.  We correct the moments
by the ratio of the value of the moment extrapolated to a common
reference value to the value of the moment for emission clipped at
$T_A=2\sigma_{rms}$ \citep{gmcs-galcen,smcgmcs}.  The correction
factor depends solely upon the peak-to-edge ratio, $P \equiv
T_{max}/T_{clip}$ where $T_{max}$ is the maximum antenna temperature
in the emission region and $T_{clip}$ is the clipping level, equal to
$2\sigma_{rms}$ for this study.  In this case, the correction factor,
$f(P)$, is
\begin{eqnarray}
f(P)&=&\left[\frac{\int_0^{x_{ref}}x^4 \exp(-x^2/2)}
{\int_0^{x_{ref}}x^2 \exp(-x^2/2)}
\right]\cdot \left[\frac{\int_0^{x_{max}}x^4 \exp(-x^2/2)}
{\int_0^{x_{max}}x^2 \exp(-x^2/2)}
\right]^{-1}\mbox{ where }\\
x_{max} &=& \sqrt{2\ln(P)} \mbox{ and } x_{ref} = 2\nonumber
\end{eqnarray}
The clouds in \citet{srby87} typically have $P=8$ implying, for that
sample, $x_{max}=\sqrt{2\ln(8)}=2.04$.  To match the clipping levels
used in that study, we choose the reference value for the
extrapolation $x_{ref}=2$.  For a typical cloud which has $P=3.5$,
$f(P) = 1.4$ resulting in a correction of 20\% to the size and
linewidth of the cloud.

The linewidth is calculated by scaling the corrected dispersion to a
FWHM using the Gaussian linewidth correction: $\Delta V =
\sigma_v\cdot \sqrt{f(P) 8\ln(2)}$.  The cloud radius is derived by
correcting $\sigma_r$ for clipping and beam convolution and then
applying the scaling from $\sigma_r$ used by \citet{srby87} in their
study of Milky Way clouds:
\begin{equation}
R_e = \frac{3.4}{\sqrt{2\pi}} \sqrt{ f(P)\sigma_r^2-2\sigma_{beam}^2}.
\end{equation}
Our simple model to correct $R_e$ for beam convolution breaks down
when $\sigma_r\sqrt{f(P)} < \sigma_{beam}$, and so we only consider a
cloud resolved if $R_e > 40\mbox{ pc}$.  Similarly, we only consider a
cloud to be resolved in the velocity direction if $\Delta V$ is larger
than the channel width.

The luminosity of the cloud is also reduced by clipping the cloud at
$2\sigma_{rms}$.  To account for this clipping, we correct the
integrated intensity value by a factor of
\begin{equation}
g(P)=\frac{\int_0^{x_{ref}}x^2 \exp(-x^2/2)}
{\int_0^{x_{max}}x^2 \exp(-x^2/2)}
\end{equation}
For clouds with $P=3.5$, the correction factor $g(P)=1.4$ for
$x_{ref}=2$.  The dominant source of uncertainty comes from the
extrapolation of cloud properties using the correction factors $f(P)$
and $g(P)$, though most of our conclusions are independent of applying
this correction.  In this paper, we have assigned the derived
quantities an uncertainty equal to half the value of this correction.


\end{document}